\newif\ifAppendix

\newcommand{\arxivVersion}{\True} 

\documentclass[runningheads]{llncs}

\usepackage{booktabs}   

\usepackage{graphicx}
\usepackage{tikz}
\usepackage{pgfplots}
\pgfplotsset{compat=1.14}
\usetikzlibrary{arrows,chains,matrix,positioning,scopes,automata,backgrounds,fit,math,plotmarks}
\usetikzlibrary{3d,shapes,fadings}
\usepackage{tabularx}
\usepackage{makecell}

\usepackage{eqnarray,amsmath}
\usepackage{amssymb}
\usepackage{enumerate}
\usepackage{bbm}
\usepackage{dsfont}
\usepackage{multirow}
\usepackage[nodayofweek]{datetime}

\usepackage[utf8]{inputenc}

\usepackage{hyperref}

\usepackage[noend]{algorithmic}
\usepackage{algorithm}

\usepackage{semantic}

\usepackage{listings}
\newcommand{\Pcode}[1]{\textsf{#1}} 

\newcommand{\localCommand}{} 
\newlength {\localLength}    

\newcommand{\newlengthsettowidth}[2]{\newlength {#1} \settowidth{#1}{#2}}
\newcommand{\newcounterset}      [2]{\newcounter{#1} \setcounter{#1}{#2}}

\newcommand{\numberRange}[1]{\mathbb #1}
\newcommand{\NN}{\numberRange N}                             

\newcommand{\range}[3][X]{\Ifthen{\Equal{#1}{X}}{\{}#2,\ldots,#3\Ifthen{\Equal{#1}{X}}{\}}} 

\newcommand{\atl}{\geq}                                 
\newcommand{\atm}{\leq}                                 
\newcommand{\st}  {\operatorname{|}}                    

\newcommand{\union}       {\mathbin        {\cup}}      
\newcommand{\Union}       {\operatorname{\bigcup}}      

\newcommand{\superseteq} {\supseteq}                    

\newcommand{\func}[3]{#1 \colon #2 \rightarrow #3}      

\newcommand{\angles}  [1]{     \langle #1       \rangle} 

\newcommand{\limplies}{\Rightarrow}               

\newcommand{\false}   {\mathit{false}}
\newcommand{\true}    {\mathit{true}}

\newcommand{\Ifthenelse}[3]{\ifthenelse{#1}{#2}{#3}}   
\newcommand{\Ifthen}    [2]{\Ifthenelse{#1}{#2}{}}
\newcommand{\Unless}    [2]{\Ifthen{\not {#1}}{#2}}
\newcommand{\Equal}     [2]{\equal{#1}{#2}}            
\newcommand{\Empty}     [1]{\Equal{#1}{}}
\newcommand{\True}         {\Equal{1}{1}}
\newcommand{\False}        {\Equal{1}{2}}

\providecommand{\Draftmode}{\False}

\newcommand{\draftcolor}{red}

\newcommand{\Itedraft}    [2]{\Ifthenelse{\Draftmode}{#1}{#2}}
\newcommand{\Ifdraft}     [1]{\Itedraft{#1}{}}

\newcommand{\drafttext}   [2][\draftcolor]{\Ifdraft{{\color{#1}#2}}}
\newcommand{\Draftpointer}{$^{*}$}
\newcommand{\draftpointer}{\makebox[0mm][c]{\Draftpointer}}
\newcommand{\draftmargin} [2][\draftcolor]{\drafttext[#1]{\draftpointer\marginpar[\raggedleft\small{\color{#1}\Draftpointer#2}]{\raggedright\small{\color{#1}\Draftpointer#2}}}} 

\newcommand{\authorcolortable}{\Ifdraft{\authorcolortablestring \end{tabular}\end{center}}}
\newcommand{\authorcolorline}[2]{{\color{#1}#1:} & {\color{#1}#2} \\}
\newcommand{\authorcolortablestring}{%
  \noindent The following colors are used in draft mode:
  \begin{center}
    \begin{tabular}{l|l}
      Color & Context \\
      \hline
      \authorcolorline{\draftcolor}{general draft text}
}

\newcommand{\draftauthor}[5]{
  \newcommand{#2}[1]{\drafttext  [#4]{##1}}  
  \newcommand{#3}[1]{\draftmargin[#4]{##1}}  
  \let#5\authorcolortablestring
  \renewcommand{\authorcolortablestring}{#5 \authorcolorline{#4}{#1}}
}

\newcommand{\ifspacecolor}{gray}
\draftauthor{ifspace}{\ifspace}{\ifspacemargin}{\ifspacecolor}{\dummyIfspace}

\Ifdraft{\setlength{\overfullrule}{10mm}}

\newcommand{\centerTwoOut}  [2]               {#1 \hfill #2}                          

\newlength{\posLength}

\newcommand{\raisetext}[2][1.3ex]{\raisebox{#1}[-#1]{#2}} 
\newcommand{\mc}{\multicolumn}                          
\newlengthsettowidth{\tabLength}{\ \ \ }

\newcommand{\wbox}[2][\tabLength]{\hspace*{#1}\mbox{#2}\hspace*{#1}}

\newcommand{\0}[1]{}                                                   
\newcommand{\End}{\end{document}}                                      
\newcommand{\format}{\drafttext{\underline{\makebox[0mm][c]{$\mid$}}}} 
\newcommand{\emphasize}[1]{\textbf{#1}}                                
\newcommand{\emphdef}[1]{\emphasize{#1}}                               

\newcommand{\defFullOrAbbrev}[2]{#1} 
\renewcommand{\defFullOrAbbrev}[2]{#2} 

\newtheorem {DEF}     {\defFullOrAbbrev{Definition} {Def.}}

\newtheorem {COR}[DEF]{\defFullOrAbbrev{Corollary}  {Cor.}}
\newtheorem {EXA}[DEF]{\defFullOrAbbrev{Example}    {Ex.}}
\newtheorem {LEM}[DEF]{\defFullOrAbbrev{Lemma}      {Lem.}}

\newtheorem {THE}[DEF]{\defFullOrAbbrev{Theorem}    {Thm.}}

\newcommand{\opNote}[2]{\stackrel{\mbox{\tiny #1}}{#2}}

\newlength{\spaceAfterEOP}
\newcommand{\Proof}{\noindent\textbf{Proof}}               
\newcommand{\eop}[1][\spaceAfterEOP]{\eopBox \vspace{#1}}  

\newcommand{\eopBox}{~\hfill$\Box$}                        

\algsetup{indent=1.0em}

\newcommand{\plkeyword}[1]{\textbf{#1}}

\newcommand{\plreturn}{\plkeyword{return}}

\newcommand{\code}[1]{\texttt{#1}}

\newcommand{\refCapitalOrSmall}[3]{#1#3} 

\newcommand{\refFullOrAbbrev}[2]{#1} 
\renewcommand{\refFullOrAbbrev}[2]{#2} 

\newcommand{\appendixref}   [2][!]{\genericref[#1] A a {\refFullOrAbbrev{ppendix}   {pp.}}  {appendix}   {#2}}
\newcommand{\algorithmref}  [2][!]{\genericref[#1] A a {\refFullOrAbbrev{lgorithm}  {lg.}}  {algorithm}  {#2}}

\newcommand{\corollaryref}  [2][!]{\genericref[#1] C c {\refFullOrAbbrev{orollary}  {or.}}  {corollary}  {#2}}
\newcommand{\definitionref} [2][!]{\genericref[#1] D d {\refFullOrAbbrev{efinition} {ef.}}  {definition} {#2}}
\newcommand{\exampleref}    [2][!]{\genericref[#1] E e {\refFullOrAbbrev{xample}    {x.}}   {example}    {#2}}
\newcommand{\figureref}     [2][!]{\genericref[#1] F f {\refFullOrAbbrev{igure}     {ig.}}  {figure}     {#2}}

\newcommand{\lemmaref}      [2][!]{\genericref[#1] L l {\refFullOrAbbrev{emma}      {em.}}  {lemma}      {#2}}
\newcommand{\lineref}       [2][!]{\genericref[#1] L l {\refFullOrAbbrev{ine}       {ine}}  {line}       {#2}}

\newcommand{\sectionref}    [2][!]{\genericref[#1] S s {\refFullOrAbbrev{ection}    {ec.}}  {section}    {#2}}
\newcommand{\tableref}      [2][!]{\genericref[#1] T t {\refFullOrAbbrev{able}      {able}} {table}      {#2}}
\newcommand{\theoremref}    [2][!]{\genericref[#1] T t {\refFullOrAbbrev{heorem}    {hm.}}  {theorem}    {#2}}

\newcommand{\equationref}[2][!]{\Ifthen{\Equal{#1}!}{\refCapitalOrSmall E e {\refFullOrAbbrev{quation}{q.}}~}(\ref{equation: #2})}

\newcommand{\genericref} [6][!]{\Ifthen{\Equal{#1}!}{\refCapitalOrSmall{#2}{#3}{#4}~}\ref{#5: #6}} 

\newcommand{\CASE}[1]{\STATE \textbf{case} #1\textbf{:} \begin{ALC@g}}
\newcommand{\ENDCASE}{\end{ALC@g}}

\newcommand{\DEFAULT}{\STATE \textbf{default:} \begin{ALC@g}}
\newcommand{\ENDDEFAULT}{\end{ALC@g}}
\newcommand{\DEFAULTLINE}[1]{\STATE \textbf{default:} }

\renewcommand{\bf}{\bfseries}

\renewcommand{\it}{\itshape}

\definecolor{mygreen}{RGB}{30, 140, 32}

\draftauthor{Thomas}{\thomas}{\thomasmargin}{mygreen}{\dummyThomas}
\draftauthor{Peizun}{\peizun}{\peizunmargin}{blue}   {\dummyPeizun}
\draftauthor{Akash} {\akash} {\akashmargin} {orange} {\dummyAkash}

\newcommand{\superscriptobject}[2]{\Ifthenelse{\Empty{#1}}{#2}{{#2}^{#1}}}              

\newcommand{\spec} {\Phi} 

\newcommand{\initindex} {I}

\newcommand{\qs}[1][] {P\Unless{\Empty{#1}}{_{#1}}} 

\newcommand{\qsstates} [1][]   {\mathcal{L}\Unless{\Empty{#1}}{_{#1}}}
\newcommand{\qsstate}[1][]   {\ell\Unless{\Empty{#1}}{_{#1}}}
\newcommand{\qsinitstate} {\superscriptobject{\initindex}{\qsstate}}

\newcommand{\alphabet}   {\Sigma}

\newcommand{\closure}[1] {\superscriptobject{*}{#1}}
\newcommand{\qstranss}[1][]   {\Delta\Unless{\Empty{#1}}{_{#1}}}

\newcommand{\transsymbol}[1][]{\Ifthenelse{\Empty{#1}}{\rightarrow}{\overset{#1}{\rightarrow}}}
\newcommand{\longtranssymbol}[1][]{\Ifthenelse{\Empty{#1}}{\longrightarrow}{\xrightarrow{#1}}}

\newcommand{\actions}[1][]   {\mathit{Act}\Unless{\Empty{#1}}{_{#1}}}
\newcommand{\action} [1][]   {\mathit{act}\Unless{\Empty{#1}}{_{#1}}}

\newcommand{\dequeue} {\mathit{deq}}
\newcommand{\noqueue} {\mathit{loc}}
\newcommand{\send   }[2] {!(#1, #2)}

\newcommand{\machineState}[1][] {m\Unless{\Empty{#1}}{_{#1}}}
\newcommand{\machineInitState} {\superscriptobject{\initindex}{\machineState}}
\newcommand{\machineConfig}[2] {({#1,#2})}
\newcommand{\systemtrans}[5] {\machineConfig {#1} {#2} \transsymbol[#3] \machineConfig {#4} {#5}}

\newcommand{\globalstate}[1][] {s\Unless{\Empty{#1}}{_{#1}}}
\newcommand{\globalconfig}[2] {\angles{#1,\ldots,#2}}

\newcommand{\queue}[1][] {\mathcal Q\Unless{\Empty{#1}}{_{#1}}} 
\newcommand{\queueAbs}[1][] {\overline{\queue}\Unless{\Empty{#1}}{_{#1}}} 
\newcommand{\emptyqueue}{\varepsilon}

\newcommand{\abstractmap}{\alpha}
\newcommand{\concretizemap}{\gamma}

\newcommand{\headmap}[1][]   {\mathcal H}

\newcommand{\abstractHeadMap}[1][]   {\abstractmap_{\tiny{\mathcal H}}}
\newcommand{\concreteHeadMap}[1][]   {\concretizemap_{\tiny{\mathcal H}}}

\newcommand  {\abstractListMap}[1][] {  \abstractmap\Unless{\Empty{#1}}{_{#1}}}
\newcommand{\concretizeListMap}[1][] {\concretizemap\Unless{\Empty{#1}}{_{#1}}}

\newcommand{\abstractqueueSep}{ \! \mid \! }
\newcommand{\abstractqueue}[3][X]{\Ifthenelse{\Empty{#1}}{#2 \abstractqueueSep #3}{\mathit{#2} \abstractqueueSep \mathit{#3}}}
\newcommand{\abstractqueueRE}[1][] {\mathit{RE}\Unless{\Empty{#1}}{_{#1}}} 

\newcommand{\lang}{\mathcal L} 

\newcommand{\emptyword}{\varepsilon} 

\newcommand{\concat}{\cdot}
\newcommand{\event}[1][]{\Ifthenelse{\Empty{#1}}{e}{e_{#1}}}

\newcommand{\concreteReach}[1][] {R\Unless{\Empty{#1}}{_{#1}}}
\newcommand{\abstractReach}[1][] {\overline{R}\Unless{\Empty{#1}}{_{#1}}}

\newcommand{\family}[2][k]{({#2})_{#1=0}^{\infty}}

\newcommand{\abstractOS}[1][k]{\family[#1]{\abstractReach[#1]}}

\newcommand{\convergencePoint} {k_{\max}}

\newcommand{\lts}[1][] {M\Unless{\Empty{#1}}{_{#1}}} 

\newcommand{\ltsStates}[1][] {S\Unless{\Empty{#1}}{_{#1}}} 
\newcommand{\ltsTranss}[1][] {T\Unless{\Empty{#1}}{_{#1}}} 
\newcommand{\ltsLabel}[1][]  {L\Unless{\Empty{#1}}{_{#1}}} 
\newcommand{\ltsPath}  {\pi} 

\newcommand{\queueIndex}[1] {\queue[][#1]}
\newcommand{\from}[2]{#1[][{\mathit{#2\rightarrow}}]}
\newcommand{\satisfies}{\models}
\newcommand{\satisfiesAbs}[1][p]{\satisfies_{#1}}

\newcommand{\temporal}[1]{\operatorname{\mathsf{#1}}}
\newcommand{\X}{\temporal X}
\newcommand{\F}{\temporal F}
\newcommand{\G}{\temporal G}

\newcommand{\numb}[1]{\#{#1}}

\newcommand{\evalf} V

\newcommand{\pfeature}[1] {\textsf{#1}}

\newcommand{\ourtool} {\textsc{Pat}}
\newcommand{\Patp} {\textsc{Pat}}
\newcommand{\Pati} {\textsc{Pat$+$I}}
\newcommand{\zing} {Zing}

\newcommand{\Ptester}{PTester} 

\usepackage{pifont}
\newcommand{\cmark}{\text{\ding{51}}}%
\newcommand{\xmark}{\text{\ding{55}}}%
\newcommand{\Yes}{\cmark}
\newcommand{\No}{\xmark}

\newcommand{\qrelop}{\triangleright}

\makeatletter
\newcommand{\bboxed}[1]{\text{\fboxsep=.02em\fcolorbox{gray}{blue!30}{\m@th$\displaystyle#1$}}}
\makeatother
\makeatletter
\newcommand{\wboxed}[1]{\text{\fboxsep=.02em\fcolorbox{gray}{gray!50}{\m@th$\displaystyle#1$}}}
\makeatother

\newcommand{\FirstMachine} {Sender} 
\newcommand{\SecondMachine}{Receiver}

\newcommand{\EventFont}[1] {\mbox{\textsc{#1}}} 
\newcommand{\eventFont}[1] {\mbox{#1}}
\newcommand{\makeEvent}[3]{
  \newcommand{#1}{\EventFont{#3}}
  \newcommand{#2}{\eventFont{#3}}
}
\makeEvent{ \FirstEvent}{ \firstEvent}{Prime}
\makeEvent{\SecondEvent}{\secondEvent}{Done}
\makeEvent{ \ThirdEvent}{ \thirdEvent}{Ping}

\newcommand{\commentcolor}{\color{black!50!green!70}}

\newcommand{\PF}{$\mathit{PiFl}$}

\newcommand{\abstractTrans}{\overline{\mathit{Im}}} 

\newcommand{\appref}[1]{App.~\textcolor{black}{#1}}

\begin{document}
\title{%
  Verifying Asynchronous Event-Driven Programs\\
  Using Partial Abstract Transformers%
  \thanks{Work supported by the US National Science Foundation under Grant
    No.~1253331, and by Microsoft Research India while hosting the second
    author for a sabbatical.}
}

\titlerunning{Verifying AED Programs Using Partial Abstract Transformers}

\author{Peizun Liu\inst{1} \and
  Thomas Wahl\inst{1} \and
Akash Lal\inst{2}}
\authorrunning{Peizun Liu \and Thomas Wahl \and Akash Lal}

\institute{Northeastern University, Boston, USA \and
Microsoft Research, Bangalore, India}

\maketitle

\begin{abstract}
  We address the problem of analyzing asynchronous event-driven programs,
  in which concurrent agents communicate via un\-bound\-ed message
  queues. The safety verification problem for such programs is
  undecidable. We present in this paper a technique that
  combines \emph{queue-bounded exploration} with a \emph{convergence test}:
  if the sequence of certain abstractions of the reachable states, for
  increasing queue bounds $k$, converges, we can prove any property of the
  program that is preserved by the abstraction. If the abstract state space
  is finite, convergence is \emph{guaranteed}; the challenge is to catch
  the point $\convergencePoint$ where it happens. We further demonstrate
  how simple invariants formulated over the \emph{concrete} domain can be
  used to eliminate spurious \emph{abstract} states, which otherwise
  prevent the sequence from converging. We have implemented our technique
  for the \pfeature P programming language for event-driven programs. We
  show experimentally that the sequence of abstractions often converges
  fully automatically, in hard cases with minimal designer support in the
  form of sequentially provable invariants, and that this happens for a
  value of $\convergencePoint$ small enough to allow the method to succeed
  in practice.
\end{abstract}

\section{Introduction}

\emph{Asynchronous event-driven (AED) programming} refers to a style of
programming multi-agent applications. The agents communicate shared work
via messages. Each agent waits for a message to arrive, and then
processes~it, possibly sending messages to other agents, in order to
collectively achieve a goal. This programming style is common for
distributed systems as well as low-level designs such as device
drivers~\cite{DGJQRZ13}. Getting such applications right is an arduous
task, due to the inherent concurrency: the programmer must defend against
all possible interleavings of messages between agents. In response to this
challenge, recent years have seen multiple approaches to verifying AED-like
programs, e.g.~by delaying send actions, or temporarily bounding their
number (to keep queue sizes~small)~\cite{DGM14,BEJQ18}, or by reasoning
about a small number of representative execution schedules, to avoid
interleaving explosion~\cite{BGKJ17}.

In this paper we consider the \pfeature P
language for AED
programming~\cite{DGJQRZ13}. A~\pfeature{P}~program consists of multiple
state machines running in parallel. Each machine has a local store, and a
message queue through which it receives events from other
machines. \pfeature P allows the programmer to formulate safety
specifications via a statement that \pfeature{assert}s some predicate over
the local state of a single machine. Verifying such reachability properties
of course requires reasoning over global system behavior and is, for
unbounded-queue \pfeature{P} programs, undecidable~\cite{BZ83}.

The unboundedness of the reachable state space does not prevent the use of
testing tools
that try to explore as much of the state space as
possible~\cite{EQR11,AAC13,DGJQRZ13,BE14}
in the quest for bugs. Somewhat inspired by this kind of approach, the goal
of this paper is a verification technique that can (sometimes) \emph{prove}
a safety property, despite exploring only a finite fraction of that
space. Our approach is as follows. Assuming that the machines' queues are
the only source of unboundedness, we consider a bound $k$ on the queue
size, and exhaustively compute the reachable states $\concreteReach[k]$ of
the resulting finite-state problem, checking the local assertion $\Phi$
along the way. We then increase the queue bound until (an error is found,
or) we reach some point $\convergencePoint$ of \emph{convergence}: a point
that allows us to conclude that increasing $k$ further is not required to
prove $\Phi$.

What kind of ``convergence'' are we targeting?
We design a sequence $\abstractOS$ of abstractions of
each reachability set over a \emph{finite} abstract state space.
Due to the monotonicity of sequence $\abstractOS$, this ensures
convergence, i.e.\ the existence of $\convergencePoint$ such that
$\abstractReach[K] = \abstractReach[\convergencePoint]$ for all $K \atl
\convergencePoint$. Provided that an abstract state satisfies $\Phi$
exactly if all its concretizations do, we have: if~all abstract states in
$\abstractReach[\convergencePoint]$ comply with $\Phi$, then so do
all reachable concrete states of
\pfeature P---we have proved the property.

We implement this strategy using an abstraction function $\abstractListMap$
with a finite co-domain that leaves the local state of a machine unchanged
and maintains the \emph{first occurrence} of each event in the queue;
repeat occurrences are dropped. This
abstraction preserves properties over the local state and the head of the
queue, i.e.\ the visible (to the machine) part of the state space, which is
typically sufficient to express reachability properties.

The second major step in our approach is the detection of the point of
convergence of $\abstractOS$: We show that, for
the \emph{best abstract transformer} $\abstractTrans$ \cite[see
  \sectionref{Abstract Convergence Detection}]{CC79,RSY04},
if $\abstractTrans(\abstractReach[k]) \subseteq \abstractReach[k]$, then
$\abstractReach[K] = \abstractReach[k]$ for all $K \atl k$. In fact, we
have a stronger result: under an easy-to-enforce condition, it suffices to
consider abstract \emph{dequeue operations}: all others, namely enqueue and
local actions, never lead to abstract states in $\abstractReach[k+1]
\setminus \abstractReach[k]$. The best abstract transformer for dequeue
actions is efficiently implementable for a given \pfeature P program.

It is of course possible that the convergence condition
$\abstractTrans(\abstractReach[k]) \subseteq \abstractReach[k]$ never holds
(the problem is undecidable). This manifests in the presence of a
\emph{spurious} abstract state in the image produced by $\abstractTrans$,
i.e.\ one whose concretization does not contain any reachable state. Our
third contribution is a technique to assist users in eliminating such
states, enhancing the chances for convergence. We have observed that
spurious abstract states are often due to violations of simple
\emph{machine invariants}: invariants that do not depend on the behavior of
other machines. By their nature, they can be proved using a cheap
sequential analysis.

We can eliminate an abstract state (e.g.~produced by $\abstractTrans$) if
\emph{all} its concretizations violate a machine invariant. In this paper,
we propose a domain-specific temporal logic to express invariants over
machines with event queues and, more importantly, an algorithm that decides
the above \emph{abstract queue invariant checking} problem, by reducing it
efficiently to a plain model checking problem. We have used this technique
to ensure the convergence in ``hard'' cases that otherwise defy convergence
of the abstract reachable states sequence.

We have implemented our technique for the \pfeature P language and
empirically evaluated it on an extensive set of benchmark programs.
The experimental results support the following conclusions: (i) for our
benchmark programs, the sequence of abstractions often converges fully
automatically, in hard cases with minimal designer support in the form of
separately dischargeable invariants; (ii) almost all examples converge at a
small value of $\convergencePoint$; and iii) the overhead our technique
adds to the bounding technique is small: the bulk is spent on the
exhaustive bounded exploration itself.

Proofs and other supporting material can be found in the
Appendix\Unless{\arxivVersion}{ of~\cite{LWL19TR}}.

\section{Overview}
\label{section: Overview}

We illustrate the main ideas of this paper using an example in the
\pfeature P language. A~machine in a \pfeature P program consists of
multiple states. Each state defines an entry code block that is executed
when the machine enters the state. The state also defines handlers for each
event type \pfeature e that it is prepared to receive. A handler can either
be \pfeature{on e do foo} (executing \pfeature{foo} on receiving \pfeature
e), or \pfeature{ignore e} (dequeuing and dropping \pfeature e). A state
can also have a \pfeature{defer e} declaration; the semantics is that a
machine dequeues the first non-\pfeature{defer}red event in its queue. As a
result, a queue in a \pfeature P program is not strictly FIFO. This
relaxation is an important feature of \pfeature P that helps programmers
express their logic compactly~\cite{DGJQRZ13}. \figureref{a running
  example} shows a \pfeature{P} program named \PF, in which a
\FirstMachine\ (eventually) floods a \SecondMachine's queue with
\ThirdEvent\ events. This queue is the only source of unboundedness in~\PF.
\begin{figure}[htbp]
  \begin{minipage}{0.45\textwidth}
\begin{lstlisting}[%frame=trBL,
      commentstyle=\commentcolor,
      framexleftmargin=2pt,        
      numbers=left,              
      % numbersep=10pt,
      numberstyle=\tiny\sffamily]
event @\FirstEvent@, @\SecondEvent@, @\ThirdEvent@;

machine @\FirstMachine@ {
  var @\secondMachine@: machine;
  start state Init {
    entry {
      @\secondMachine@ = new @\SecondMachine@();
    }
    goto @\firstEvent@_it;
  }
  state @\firstEvent@_it {
    entry {
      var i:int;
      while (i < 3) { // 3x @{\commentcolor\FirstEvent}@
        send @\secondMachine@, @\FirstEvent@; i = i + 1;
      }
      send @\secondMachine@, @\SecondEvent@; goto @\thirdEvent@_it;
    }
  }
\end{lstlisting}
  \end{minipage}
  \hfill
  \begin{minipage}{0.45\textwidth}
    \vskip8pt
\begin{lstlisting}[%frame=trBL,
      firstnumber=20,
      numbers=left,
      numberstyle=\tiny\sffamily,
      commentstyle=\commentcolor,
      morekeywords={set}, 
      framexleftmargin=2pt]
  state @\thirdEvent@_it {
    entry {
      send @\secondMachine@, @\ThirdEvent@; goto @\thirdEvent@_it; 
    }
  }
}

machine @\SecondMachine@ {
  start state Init {
    defer @\FirstEvent@;
    on @\SecondEvent@ goto Ignore_it; 
  }

  state Ignore_it { 
    ignore @\FirstEvent, \ThirdEvent@;
  }
}
\end{lstlisting}
  \end{minipage}
  \caption{\PF: a \thirdEvent-Flood scenario. The \FirstMachine\ and the
    \SecondMachine\ communicate via events of types \FirstEvent,
    \SecondEvent, and \ThirdEvent. After sending some \FirstEvent\ events
    and one \SecondEvent, the \FirstMachine\ floods the
    \SecondMachine\ with \ThirdEvent s. The \SecondMachine\ initially
    \pfeature{defer}s \FirstEvent s. Upon receiving \SecondEvent\ it enters
    a state in which it \pfeature{ignore}s \ThirdEvent.
%
  }
  \label{figure: a running example}
\end{figure}

A critical property for \pfeature P programs is \emph{(bounded)
  responsiveness}: the receiving machine must have a handler
(e.g.~\pfeature{on}, \pfeature{defer}, \pfeature{ignore}) for every event
arriving at the queue head; otherwise the event will come as a ``surprise''
and crash the machine. To prove responsiveness for \PF, we have to
demonstrate (among others) that in state {\sffamily Ignore\_it}, the head
of the \SecondMachine's queue is always \ThirdEvent. We cannot perform
exhaustive model checking, since the set of reachable
states is infinite. Instead, we will compute a conservative abstraction of
this set that is precise enough to rule out \FirstEvent\ or
\SecondEvent\ events at the queue head in this state.

We first define a suitable abstraction function $\abstractListMap$ that
collapses repeated occurrences of events to each event's first
occurrence. For instance, the queue
\begin{equation}
  \label{equation: example queue}
  \queue = \mbox{\FirstEvent.\FirstEvent.\FirstEvent.\SecondEvent.\ThirdEvent.\ThirdEvent.\ThirdEvent.\ThirdEvent}
\end{equation}
will be abstracted to $\queueAbs = \abstractListMap(\queue) =
\mbox{\FirstEvent.\SecondEvent.\ThirdEvent}$. The \emph{finite} number of
possible abstract queues is $1 + 3 + 3 \cdot 2 + 3 \cdot 2 \cdot 1 =
16$. The abstraction preserves the head of the queue. This and the machine
state is enough information to check responsiveness.

We now generate the sequence $\abstractReach[k]$ of abstractions of the
reachable states sets $\concreteReach[k]$ for queue size bounds
$k=0,1,2,\ldots$, by computing each finite set $\concreteReach[k]$, and
then $\abstractReach[k]$ as~$\abstractListMap(\concreteReach[k])$. The
obtained monotone sequence $\abstractOS$ over a finite domain will
eventually converge, but we must prove that it has. This is done by
applying the \emph{best abstract transformer} $\abstractTrans$, restricted
to dequeue operations (defined in \sectionref{Abstract Convergence
  Detection}), to the current set $\abstractReach[k]$, and confirming that
the result is contained in $\abstractReach[k]$.

As it turns out, the confirmation fails for the \PF\ program: $k=5$ marks
the first time set $\abstractReach[k]$ repeats, i.e.\ $\abstractReach[4] =
\abstractReach[5]$, so we are motivated to run the convergence
test. Unfortunately we find a state $\bar s \in
\abstractTrans(\abstractReach[5]) \setminus \abstractReach[5]$,
preventing convergence. Our approach now offers two remedies to this
dilemma. One is to refine the queue abstraction. In our implementation,
function $\abstractListMap$ is really $\abstractListMap[p]$, for a
parameter $p$ that denotes the size of the \emph{prefix} of the queue that
is kept unchanged by the abstraction. For example, for the queue from
\equationref{example queue} we have $\abstractListMap[4](\queue) =
\mbox{\FirstEvent.\FirstEvent.\FirstEvent.\SecondEvent} \mid
\mbox{\ThirdEvent}$, where $\mid$ separates the prefix from the ``infinite
tail'' of the abstract queue. This (straightforward) refinement maintains
finiteness of the abstraction and increases precision, by revealing that
the queue starts with three $\FirstEvent$ events. Re-running the analysis for the \PF\ program with $p=4$, at
$k=5$ we find $\abstractTrans(\abstractReach[5]) \subseteq
\abstractReach[5]$, and the proof is complete.

The second remedy to the failed convergence test dilemma is more powerful
but also less automatic. Let's revert to prefix $p=0$ and inspect the
abstract state $\bar s \in \abstractTrans(\abstractReach[5]) \setminus
\abstractReach[5]$ that foils the test. We find that it features a
\SecondEvent\ event followed by a \FirstEvent\ event in the
\SecondMachine's queue. A simple static analysis of the \FirstMachine's
machine in isolation shows that it permits no path from the
\mbox{\Pcode{send \SecondEvent}} to the \mbox{\Pcode{send \FirstEvent}}
statement. The behavior of other machines is irrelevant for this invariant;
we call it a \emph{machine invariant}. We pass the invariant to our tool
via the command line using the expression
\begin{equation}
  \label{equation: motivating example QuTL}
  \G \, (\SecondEvent \limplies \G \lnot \FirstEvent)
\end{equation}
in a temporal-logic like notation called QuTL (\sectionref{Queue Temporal
  Logic (QuTL)}), where $\G$ universally quantifies over all queue
entries. Our tool includes a QuTL checker that determines that
\emphasize{every concretization} of $\bar s$ violates property
\equationref[]{motivating example QuTL}, concluding that $\bar s$ is
spurious and can be discarded. This turns out to be sufficient for
convergence.

\section{Queue-(Un)Bounded Reachability Analysis}
\label{section: P Program Verification via Sequence Convergence}

\subsubsection{Communicating Queue Systems.}

We consider \pfeature P programs consisting of a fixed and known number $n$
of machines communicating via event passing through unbounded FIFO
queues.\footnote{The \pfeature P language permits unbounded machine
  creation, a feature that we do not allow here and that is not used in any
  of the benchmarks we are aware of.} For simplicity, we assume the
machines are created at the start of the program; dynamic creation at a
later time can be simulated by having the machine \pfeature{ignore} all
events until it receives a special creation event.

We model such a program as a \emph{communicating queue system}
(CQS). Formally, given $n \in \NN$, a CQS $\qs^n$ is a collection of $n$
\emph{queue automata} (QA) $\qs[i] =
(\alphabet,\qsstates[i],\actions[i],\qstranss[i],\qsstate[i]^I)$, $1 \atm i
\atm n$. A QA consists of a finite queue alphabet $\alphabet$ shared by all
QA, a finite set $\qsstates[i]$ of local states, a finite set $\actions[i]$
of action labels, a finite set $\qstranss[i] \subseteq \qsstates[i] \times
(\alphabet \union \{ \emptyqueue \}) \times \actions[i] \times \qsstates[i]
\times (\alphabet \union \{ \emptyqueue \})$ of transitions, and an initial
local state $\qsstate[i]^I \in \qsstates[i]$. An action label $\action \in
\actions[i]$
is of the form
\begin{itemize}

\item $\action \in \{\dequeue,\noqueue\}$, denoting an action \emph{internal}
  to $\qs[i]$ (no other QA involved) that either \emph{deq}ueues an event
  ($\dequeue$), or updates its \emph{loc}al state
  ($\noqueue$); \hfill \emphasize{or}

\item $\action = \ \send{\event}{j}$, for $\event \in \alphabet$, $j \in
  \range 1 n$, denoting a \emph{transmission}, where $\qs[i]$
  (the \emph{sender}) adds event $\event$ to the end of the queue of
  $\qs[j]$ (the \emph{receiver}).
\end{itemize}

The individual QA of a CQS model machines of a \pfeature{P} program; hence
we refer to QA states as \emph{machine states}. A~transmit action is the
only communication mechanism among the QA.

\paragraph{Semantics.} A \emph{machine state} $\machineState$ of a QA is of
the form $\machineConfig \qsstate \queue \in \qsstates \times \closure{\alphabet}$;
state $\machineInitState = \machineConfig{\qsinitstate}{\emptyqueue}$ is
\emph{initial}.
We define machine transitions corresponding to internal actions as follows
(transmit actions are defined later at the global level):
\[
  \begin{array}{lcl}
  \inference{ %
    \systemtrans {\qsstate} {\emptyqueue} {\noqueue} {\qsstate'} {\emptyqueue} \in \qstranss }
  {\machineConfig \qsstate \queue \transsymbol \machineConfig {\qsstate'} {\queue}}
  [for $\qsstate,\qsstate' \in \qsstates$, $\queue \in \closure{\alphabet}$]                         & \quad \quad & \mbox{(\bf local)}\\[5mm]
  \inference{ %
   \systemtrans {\qsstate} {\event} {\dequeue} {\qsstate'} {\emptyqueue} \in \qstranss }
  {\machineConfig \qsstate {\event\queue} \transsymbol \machineConfig {\qsstate'} {\queue}}
  [for $\qsstate,\qsstate' \in \qsstates$, $\event \in \alphabet$, $\queue \in \closure{\alphabet}$] &             & \mbox{(\bf dequeue)}
  \end{array}
\]

A \emph{(global) state} $\globalstate$ of a CQS is a tuple $\globalconfig
{\machineConfig {\qsstate[1]} {\queue[1]}} {\machineConfig {\qsstate[n]}
  {\queue[n]}}$ where ${\machineConfig {\qsstate[i]} {\queue[i]}} \in
\qsstates[i] \times \closure{\alphabet}$ for $i \in \range 1 n$.
State $\globalstate^I =$ $\globalconfig {\machineConfig {\qsstate[1]^{I}}
  {\emptyqueue}} {\machineConfig {\qsstate[n]^{I}} {\emptyqueue}}$ is
initial. We extend the machine transition relation $\transsymbol$ to states
as follows:
\[
  \globalconfig {\machineConfig {\qsstate[1]} {\queue[1]}} {\machineConfig {\qsstate[n]} {\queue[n]}}
  \transsymbol
  \globalconfig {\machineConfig {\qsstate[1]'} {\queue[1]'}} {\machineConfig {\qsstate[n]'} {\queue[n]'}}
\]
if there exists $i \in \range 1 n$ such that one of the following holds:
\begin{description}

\item[(internal)] $\machineConfig {\qsstate[i]} {\queue[i]} \transsymbol
  \machineConfig {\qsstate[i]'} {\queue[i]'}$, and for all $k \in \range 1
  n \setminus \{i\}$, $\qsstate[k] = \qsstate[k]'$, $\queue[k] =
  \queue[k]'$;\vspace{2mm}

\item[(transmission)] there exists $j \in \range 1 n$ and $\event \in \alphabet$ such that:
  \begin{enumerate}

  \item $\machineConfig {\qsstate[i]} {\emptyqueue} \longtranssymbol[\send{\event}{j}] \machineConfig {\qsstate[i]'}{\emptyqueue} \in \qstranss[i]$ ;

  \item $\queue[j]' = \queue[j]\event$ ;

  \item $\qsstate[k]'= \qsstate[k]$ for all $k \in \range 1 n \setminus \{i\}$ ; and

  \item $\queue[k]' = \queue[k]$ for all $k \in \range 1 n \setminus \{j\}$ .

  \end{enumerate}

\end{description}
The execution model of a CQS is strictly interleaving. That is, in each
step, one of the two above transitions \textbf{(internal)} or
\textbf{(transmission)} is performed for a nondeterministically chosen
machine $i$.

\subsubsection{Queue-bounded and queue-unbounded reachability.}
Given a CQS $\qs^n$, a state $\globalstate =
\globalconfig{\machineConfig{\qsstate[1]}{\queue[1]}}{\machineConfig{\qsstate[n]}{\queue[n]}}$,
and a number $k$, the \emph{queue-bounded reachability problem} (for
$\globalstate$ and $k$) determines whether $\globalstate$ is
\emph{reachable under queue bound $k$}, i.e.\ whether there exists a path
$\globalstate[0] \transsymbol \globalstate[1] \ldots \transsymbol
\globalstate[z]$ such that $\globalstate[0] = \globalstate^I$,
$\globalstate[z] = \globalstate$, and for $i \in \range 0 z$, all queues in
state $\globalstate[i]$ have at most $k$ events. Queue-bounded reachability
for $k$ is trivially decidable, by making enqueue actions for queues of
size $k$ \emph{blocking} (the sender cannot continue), which results in a
finite state space. We write $\concreteReach[k] = \{\globalstate:
\mbox{$\globalstate$ is reachable under queue bound $k$}\}$.

Queue-bounded reachability will be used in this paper as a tool for solving
our actual problem of interest: Given a CQS $\qs^n$ and a state
$\globalstate$, the \emph{Queue-UnBounded reachability Analysis (QUBA)
  problem} determines whether $\globalstate$ is reachable, i.e.\ whether
there exists a (queue-unbounded) path from $\globalstate^I$ to
$\globalstate$. The QUBA problem is undecidable~\cite{BZ83}. We write
$\concreteReach$ ($= \Union_{k \in \NN} \concreteReach[k]$) for the set of
reachable~states.

\section{Convergence via Partial Abstract Transformers}
\label{section: Abstract Convergence}

In this section, we formalize our approach to detecting the convergence of
a suitable sequence of \emph{observations} about the states
$\concreteReach[k]$ reachable under $k$-bounded semantics. We define the
observations as abstractions of those states, resulting in sets
$\abstractReach[k]$.
We then investigate the convergence of the sequence $\abstractOS$.

\subsection{List Abstractions of Queues}
\label{section: List Abstractions of Queues}

Our abstraction function applies to queues, as defined below. Its action on
machine and system states then follows from the hierarchical design of a
CQS. Let $|\queue|$ denote the number of events in $\queue$, and
$\queueIndex i$ the $i$th event in $\queue$ ($0 \atm i <
|\queue|$).
\begin{DEF}
  \label{definition: list abstraction}
  For a parameter $p \in \NN$, the \emphasize{list abstraction} function
  $\abstractListMap[p]: {\closure{\alphabet}} \mapsto {\closure{\alphabet}}$
  is defined as follows:
  \begin{enumerate}

  \item $\abstractListMap[p](\emptyqueue) = \emptyqueue$.

  \item For a non-empty queue $\queue = \mathcal P \concat \event$,
    \begin{equation}
      \label{equation: listmap}
      \abstractListMap[p](\queue) =
      \left\{
      \begin{array}{ll}
        \abstractListMap[p](\mathcal P)                & \ \ \mbox{if there exists $j$ s.t.\ $p \atm j < |\mathcal P|$ and $\queueIndex{j} = \event$} \\
        \abstractListMap[p](\mathcal P) \concat \event & \ \ \mbox{otherwise}
      \end{array}
      \right. \ .
    \end{equation}
  \end{enumerate}
\end{DEF}
Intuitively, $\abstractListMap[p]$ abstracts a queue by leaving its first
$p$ events unchanged (an idea also used in \cite{GJJ06}). Starting from
position $p$ it keeps only the first occurrence of each event $\event$ in
the queue, if any; repeat occurrences are dropped.\footnote{Note that the
  head of the queue is always preserved by $\abstractListMap[p]$, even for
  $p=0$.} The preservation of existence and order of the first occurrences
of all present events motivates the term \emph{list abstraction}. An
alternative is an abstraction that keeps only the \emph{set} (not: list) of
queue elements from position~$p$, i.e.\ it ignores multiplicity \emph{and
  order}. This is by definition less precise than the list abstraction and
provided no efficiency advantages in our experiments. An abstraction that
keeps only the queue head proved cheap but too imprecise.

The motivation for parameter $p$ is that many protocols proceed in
\emph{rounds} of repeating communication patterns, involving a bounded
number of message exchanges. If $p$ exceeds that number, the list
abstraction's loss of information may be immaterial.

We write an abstract queue $\queueAbs = \abstractListMap[p](\queue)$ in the
form $\abstractqueue{pref}{suff}$ s.t.\ $p=|\mathit{pref}|$, and refer to
$\mathit{pref}$ as $\queueAbs$'s \emph{prefix} (shared with $\queue$), and
$\mathit{suff}$ as $\queueAbs$'s \emph{suffix}.
\begin{EXA}
  \label{example: list abstraction}
  The queues $\queue \in \{\mathit{bbbba},\mathit{bbba},\mathit{bbbaa}\}$
  are $\abstractListMap[2]$-\emphdef{equivalent}:
  $\abstractListMap[2](\queue) = \abstractqueue{bb}{ba}$.
\end{EXA}


We extend $\abstractListMap[p]$ to act on a machine state via
$\abstractListMap[p]{\machineConfig{\qsstate[i]}{\queue[i]}} =
\machineConfig{\qsstate[i]}{\abstractListMap[p](\queue[i])}$, on a state
via $\abstractListMap[p](\globalstate) = $ ${\globalconfig {\machineConfig
    {\qsstate[1]} {\abstractListMap[p](\queue[1])}} {\machineConfig
    {\qsstate[n]} {\abstractListMap[p](\queue[n])}}}$, and on a set of
states pointwise via $\abstractListMap[p](S) = \{
\abstractListMap[p](\globalstate): \globalstate \in S \}$.

\paragraph{Discussion.} The abstract state space is finite since the queue
prefix is of fixed size, and each event in the suffix is recorded at most
once (the event alphabet is finite). The sets of reachable abstract states
grow monotonously with increasing queue size bound $k$, since the sets of
reachable concrete states do:
\[
  k_1 \atm k_2 \wbox{$\limplies$} R_{k_1} \subseteq R_{k_2} \wbox{$\limplies$} \abstractListMap[p](R_{k_1}) \subseteq \abstractListMap[p](R_{k_2}) \ .
\]
Finiteness and monotonicity guarantee convergence of the sequence of
reachable abstract states.

We say the abstraction function $\abstractListMap[p]$ \emph{respects} a
property of a state if, for any two $\abstractListMap[p]$-equivalent states
(see \exampleref{list abstraction}),
the property holds for both or for neither. Function $\abstractListMap[p]$
respects properties that refer to the local-state part of a machine, and to
the first $p+1$ events of its queue (which are preserved by
$\abstractListMap[p]$). In addition, the property may look beyond the
prefix and refer to the existence of events in the queue, but not their
frequency or their order after the first occurrence.

The rich information preserved by the abstraction (despite being
finite-state) especially pays off in connection with the \pfeature{defer}
feature in the \pfeature{P} language, which allows machines to delay
handling certain events at the head of a queue~\cite{DGJQRZ13}. The machine
identifies the first non-deferred event in the queue, a piece of
information that is precisely preserved by the list abstraction (no matter
what $p$).
\begin{DEF}
  \label{definition: concretizemap}
  Given an abstract queue $\queueAbs = \abstractqueue[]{\event[0] \ldots
    \event[p-1]}{\event[p] \ldots \event[z-1]}$, the
  \emphdef{concretization function}
  $\func{\concretizeListMap[p]}{\closure{\alphabet}}{2^{\closure{\alphabet}}}$
  maps $\queueAbs$ to the \emph{language} of the regular expression
  \begin{equation}
    \label{equation: concretizemap regexp}
    \abstractqueueRE[p] (\queueAbs) :=
    \event[0] \ldots \event[p-1]
    \event[p] \closure{\{ \event[p] \}}
    \event[p+1] \closure{\{ \event[p],\event[p+1]\}} 
    \ldots 
    \event[z-1] \closure{\{ \event[p], \ldots, \event[z-1] \}} \ ,
  \end{equation}
  i.e.\ $\concretizeListMap[p](\queueAbs) :=
  \lang(\abstractqueueRE[p](\queueAbs))$.
\end{DEF}
As a special case, $\abstractqueueRE[p](\emptyqueue) = \emptyword$ and so
$\concretizeListMap[p](\emptyqueue) = \lang(\emptyword) = \{\emptyword\}$
for the empty queue. We extend $\concretizeListMap[p]$ to act on abstract
(machine or global) states in a way analogous to the extension of
$\abstractListMap[p]$, by moving it inside to the queues occurring in those
states.

\subsection{Abstract Convergence Detection}
\label{section: Abstract Convergence Detection}

Recall that finiteness and monotonicity of the sequence $\abstractOS$
guarantee its convergence, so nothing seems more suggestive than to compute
the limit. We summarize our overall procedure to do so in
\algorithmref{Verifying P using abstracted observation sequences}. The
procedure iteratively increases the queue bound $k$ and computes the
concrete and (per $\abstractListMap[p]$-projection) the abstract
reachability sets $\concreteReach[k]$ and $\abstractReach[k]$. If, for some
$k$, an error is detected, the procedure terminates (Lines~\lineref[]{Ck
  violates spec}--\lineref[]{unsafe}; in practice implemented as an
on-the-fly~check).
\vspace{-1em}
\newcommand{\dequeuetranssymbol}{\stackrel{\dequeue}{\transsymbol}}
\newcommand{\enqueuetranssymbol}{\stackrel{!}{\transsymbol}}
\newcommand{\noqueuetranssymbol}{\stackrel{\noqueue}{\transsymbol}}
\newcommand{\batsymbol}{\overline T} 
\begin{algorithm}[htbp]
  \floatname{algorithm}{Algorithm}
  \begin{algorithmic}[1]
    \REQUIRE CQS with transition relation $\transsymbol$ , $p \in \NN$, property $\spec$ respected by $\abstractListMap[p]$.
    \STMT \plkeyword{compute} $\concreteReach[0]$; \ \code{$\abstractReach[0]$ := $\abstractListMap[p](\concreteReach[0])$}    \label{line: concrete and abstract R}
    \FOR {\code{$k$ := 1 to $\infty$}}                                                                                         \label{line: main loop}
      \STMT \plkeyword{compute} $\concreteReach[k]$; \ \code{$\abstractReach[k]$ := $\abstractListMap[p](\concreteReach[k])$}  \label{line: build Ak}
      \IF {$\exists r \in \concreteReach[k] : r \not\satisfies \spec$}                                                         \label{line: Ck violates spec}
        \STMT \plreturn\ ``error reachable with queue bound $k$''                                                              \label{line: unsafe}
      \ENDIF
      \IF{\code{$|\abstractReach[k]|$ = $|\abstractReach[k-1]|$}}                                                              \label{line: Ak=Ak-1}
        \STMT \code{$\batsymbol$ := $(\abstractListMap[p] \circ \mathit{Im}_{\dequeue} \circ \concretizeListMap[p])(\abstractReach[k])$} \label{line: best abstract transformer} \COMMENT{\emphasize{\emph{partial} best abstract transformer}}
        \IF{$\batsymbol \subseteq \abstractReach[k]$}                                                       \label{line: if converged}
          \STMT \plreturn\ ``safe for any queue bound''                                                                        \label{line: safe}
        \ENDIF
      \ENDIF
    \ENDFOR
  \end{algorithmic}
  \caption{Queue-unbounded reachability analysis}
  \label{algorithm: Verifying P using abstracted observation sequences}
\end{algorithm}

The key of the algorithm is reflected in
Lines~\lineref[]{Ak=Ak-1}--\lineref[]{safe} and is based on the following
idea (all claims are proved as part of \theoremref{convergence theorem}
below). If the computation of $\abstractReach[k]$ reveals no new abstract
states in round $k$ (\lineref{Ak=Ak-1}; by monotonicity, ``same size''
implies ``same sets''), we apply the \emph{best abstract transformer}
\cite{CC79,RSY04} $\abstractTrans := \abstractListMap[p] \circ
\mathit{Im}_{\transsymbol} \circ \concretizeListMap[p]$ to
$\abstractReach[k]$: if the result is contained in $\abstractReach[k]$, the
abstract reachability sequence has converged. However, we can do better: we
can restrict the successor function $\mathit{Im}_{\transsymbol}$ of the CQS
to \emph{dequeue} actions, denoted $\mathit{Im}_{\dequeue}$ in
\lineref{best abstract transformer}. The ultimate reason is that firing a
local or transmit action on two $\abstractListMap[p]$-equivalent states $r$
and $s$ results again in $\abstractListMap[p]$-equivalent states $r'$ and
$s'$. This fact does \emph{not} hold for dequeue actions: the successors
$r'$ and $s'$ of dequeues depend on the abstracted parts of $r$ and $s$,
resp., which may differ and become ``visible'' during the dequeue
(e.g.\ the event behind the queue head moves into the head position). Our
main result therefore is: if $\abstractReach[k] = \abstractReach[k-1]$ and
dequeue actions do not create new abstract states (Lines~\lineref[]{best
  abstract transformer} and~\lineref[]{if converged}), sequence
$\abstractOS$ has converged: \newcounterset{convergenceTHE}{\theDEF}
\begin{THE}
  \label{theorem: convergence theorem}
  If $\abstractReach[k] = \abstractReach[k-1]$ and $\batsymbol \subseteq
  \abstractReach[k]$, then for any $K \atl k$, $\abstractReach[K] =
  \abstractReach[k]$.
\end{THE}

If the sequence of reachable abstract states has converged, then
\emphasize{all} reachable concrete states (any $k$) belong to
$\concretizeListMap[p](\abstractReach[k])$ (for the current $k$). Since the
abstraction function $\abstractListMap[p]$ respects property $\spec$, we
know that if any reachable concrete state violated $\spec$, so would any
other concrete state that maps to the same abstraction. However, for each
abstract state in $\abstractReach[k]$,
\lineref{Ck violates spec} has examined at least one state $r$ in its
concretization; a violation was not found. We conclude:
\begin{COR}
  \label{corollary: partial correctness when converged}
  \lineref{safe} of \algorithmref{Verifying P using abstracted observation
    sequences} correctly asserts that no reachable concrete state of the
  given CQS violates $\spec$.
\end{COR}

The corollary (along with the earlier statement about Lines~\lineref[]{Ck
  violates spec}--\lineref[]{unsafe}) confirms the partial correctness of
\algorithmref{Verifying P using abstracted observation sequences}. The
procedure is, however, necessarily incomplete:
if no error is detected and the convergence condition in \lineref{if
converged} never holds, the \plkeyword{for} loop will run forever.

We conclude this part with two comments. First, note that we do not compute
the sets $\abstractReach[k]$ as reachability fixpoints in the abstract
domain (i.e.~the domain of~$\abstractListMap[p]$). Instead, we compute the
\emph{concrete} reachability sets first, and then obtain the
$\abstractReach[k]$ via projection (\lineref{concrete and abstract R}). The
reason is that the projection gives us the \emph{exact} set of abstractions
of reachable concrete states, while an abstract fixpoint likely
overapproximates (for instance, the best abstract transformer from
\lineref{best abstract transformer} does) and loses precision. Note that a
primary motivation for computing abstract fixpoints, namely that the
concrete fixpoint may not be computable, does not apply here: the concrete
domains are finite, for each $k$.

Second, we observe that this projection technique comes with a cost:
sequence $\abstractOS$ may \emph{stutter} at intermediate moments:
$\abstractReach[k] \subsetneq \abstractReach[k+1] = \abstractReach[k+2]
\subsetneq\abstractReach[k+3]$. The reason is that $\abstractReach[k+3]$ is
not obtained as a functional image of $\abstractReach[k+2]$,
but by projection from $\concreteReach[k+3]$. As a consequence, we cannot
short-cut the convergence detection by just ``waiting'' for $\abstractOS$
to stabilize, despite the finite domain.

\subsection{Computing Partial Best Abstract Transformers}
\label{section: Computing Best Abstract Transformers}

Recall that in \lineref{best abstract transformer} we compute
\begin{equation}
  \label{equation: partial RkbarPrime}
  \batsymbol = \abstractTrans_{\dequeue}(\abstractReach[k]) = (\abstractListMap[p] \circ \mathit{Im}_{\dequeue} \circ \concretizeListMap[p])(\abstractReach[k]) \ .
\end{equation}
The line applies the best abstract transformer, restricted to dequeue
actions, to~$\abstractReach[k]$. This result cannot be computed as defined
in \equationref[]{partial RkbarPrime}, since
$\concretizeListMap[p](\abstractReach[k])$ is typically infinite. However,
$\abstractReach[k]$ is finite, so we can iterative over $\bar r \in
\abstractReach[k]$, and little information is actually needed to determine
the abstract successors of $\bar r$. The ``infinite fragment'' of $\bar r$
remains unchanged, which makes the action implementable.

Formally, let $\bar r = \machineConfig{\qsstate}{\queueAbs}$ with
$\queueAbs =
\abstractqueue[]{\event[0]\event[1]\ldots\event[p-1]}{\event[p]\event[p+1]\ldots\event[z-1]}$. To
apply a dequeue action to $\bar r$, we first perform local-state updates on
$\qsstate$ as required by the action, resulting in $\qsstate'$. Now
consider $\queueAbs$. The first suffix event, $\event[p]$, moves into the
prefix due to the dequeue. We do not know whether there are later
occurrences of $\event[p]$ before or after the first suffix occurrences of
$\event[p+1] \ldots \event[z-1]$. This information determines the possible
abstract queues resulting from the dequeue. To compute the exact best
abstract transformer, we enumerate these~possibilities:
\newcommand{\ep}{\fbox{$\event[p]$}}
\begin{multline*}
  \abstractTrans_{\dequeue}
  (\{\machineConfig{\qsstate}{\queueAbs}\}) \wbox{$=$} \\[-15pt]
  \{ \ \machineConfig{\qsstate'}{\queueAbs'} \ : \ \queueAbs' \in
  \left\{
  \begin{array}{l}
    \event[1]\ldots\event[p] \abstractqueueSep \event[p+1]\event[p+2]\ldots\event[z-1]    \\
    \event[1]\ldots\event[p] \abstractqueueSep \ep\event[p+1]\event[p+2]\ldots\event[z-1] \\
    \event[1]\ldots\event[p] \abstractqueueSep \event[p+1]\ep\event[p+2]\ldots\event[z-1] \\
    \mc{1}{c}{\vdots} \\
    \event[1]\ldots\event[p] \abstractqueueSep \event[p+1]\event[p+2]\ldots\event[z-1]\ep
  \end{array}
  \right\}
  \ \}
\end{multline*}
The first case for $\queueAbs'$ applies if there are no occurrences of
$\event[p]$ in the suffix after the dequeue. The remaining cases enumerate
possible positions of the \emph{first} occurrence of $\event[p]$ (boxed,
for readability) in the suffix after the dequeue. The cost of this
enumeration is linear in the length of the suffix of the abstract
queue.

Since our list abstraction maintains the first occurrence of each event,
the semantics of \pfeature{defer} (see the \emph{Discussion} in
\sectionref{List Abstractions of Queues}) can be implemented abstractly
without loss of information (not shown above, for simplicity).

\section{Abstract Queue Invariant Checking}
\label{section: Abstract Queue Invariant Checking}

The abstract transformer function in \sectionref{Abstract Convergence} is
used to decide whether sequence $\abstractOS$ has converged. Being an
overapproximation, the function may generate \emph{spurious} states: they
are not reachable, i.e.\ no concretization of them~is. Unfortunate for us,
spurious abstract states always prevent convergence.

A key empirical observation is that concretizations of spurious abstract
states often violate simple machine invariants, which can be
proved from the
perspective of a single machine, while collapsing all other machines into a
nondeterministically behaving environment.
Consider our example from \sectionref{Overview} for $p=0$. It fails to
converge since \lineref{best abstract transformer} generates an abstract
state $\bar s$ that features a \SecondEvent\ event followed by a
\FirstEvent\ event in the \SecondMachine's queue. A light-weight static
analysis proves that the \FirstMachine's machine permits no path from the
\mbox{\Pcode{send \SecondEvent}} to the \mbox{\Pcode{send \FirstEvent}}
statement.
Since \emphasize{every} concretization of $\bar s$ features a
\Pcode{\SecondEvent} followed by a \Pcode{\FirstEvent} event, the abstract
state $\bar s$ is spurious and can be eliminated.
 
Our tool assists users in \emph{discovering} candidate machine invariants,
by facilitating the inspection of states in $\batsymbol \setminus
\abstractReach[k]$ (which foil the test in \lineref{if converged}). We
\emph{discharge} such invariants separately, via a simple sequential
model-check or static analysis. In the section we focus on the more
interesting question of how to \emph{use} them.
Formally, suppose the \pfeature P program comes with a \emph{queue
  invariant} $I$, i.e.\ an invariant property of \emph{concrete}
queues. The \emph{abstract invariant checking problem} is to decide, for a
given abstract queue $\queueAbs$, whether \emph{every} concretization of
$\queueAbs$ violates $I$; in this case, and this case only, an abstract
state containing $\queueAbs$ can be eliminated. In the following we define
a language QuTL for specifying concrete queue
invariants~(\sectionref[]{Queue Temporal Logic (QuTL)}), and then show how
checking an abstract queue against a QuTL invariant can be efficiently
solved as a model checking problem~(\sectionref[]{Abstract QuTL Model
  Checking}).

\subsection{Queue Temporal Logic (QuTL)}
\label{section: Queue Temporal Logic (QuTL)}

Our logic to express invariant properties of queues is a form of
first-order linear-time temporal logic. This choice is motivated by the
logic's ability to constrain the order (via temporal operators) and
multiplicity of queue events, the latter via relational operators that
express conditions on the number of event occurrences.

\paragraph{Queue Relational Expressions (QuRelE).} These are of the form
$\numb e \qrelop c$, where $e \in \alphabet$ (queue alphabet), $\qrelop \in
\{<,\atm,=,\atl,>\}$, and $c \in \NN$ is a literal natural number. The
\emph{value} of a QuRelE is defined as the Boolean
\begin{equation}
  \label{equation: queue relational expression: semantics}
  \evalf(\numb e \qrelop c) \wbox{$=$} |\{i \in \NN: 0 \atm i < |\queue| \land \queueIndex{i} = \event \}| \ \qrelop \ c
\end{equation}
where $|\cdot|$ denotes set cardinality and $\qrelop$ is interpreted as the
standard integer arithmetic relational operator.
In the following we write $\from \queue i$ (read:
``$\queue$~from~$i$'') for the queue obtained from queue $\queue$ by
dropping the first $i$ events.
\begin{DEF}[Syntax of QuTL]
  \label{definition: QuTL syntax}
  The following are QuTL formulas:
  \begin{itemize}

  \item $\false$ and $\true$.

  \item $\event$, for $\event \in \alphabet$.

  \item $E$, for a queue relational expression $E$.

  \item $\X \phi$, $\F \phi$, $\G \phi$, for a QuTL formula $\phi$.

  \end{itemize}
  The set QuTL is the Boolean closure of the above set of formulas.
\end{DEF}
\begin{DEF}[Concrete semantics of QuTL]
  \label{definition: QuTL concrete semantics}
  Concrete queue $\queue$ \emphdef{satisfies} QuTL formula~$\phi$, written
  $\queue \satisfies \phi$, depending on the form of~$\phi$ as follows.
  \begin{itemize}

  \item $\queue \satisfies \true$.

  \item for $\event \in \alphabet$, $\queue \satisfies \event$ iff
    $|\queue| > 0$ and $\queueIndex{0} = \event$.

  \item for a queue relational expression $E$, $\queue \satisfies E$ iff $V(E) = \true$.

  \item $\queue \satisfies \X \phi$ iff $|\queue| > 0$ and $\from \queue 1 \satisfies \phi$.

  \item $\queue \satisfies \F \phi$ iff there exists $i \in \NN$ such that $0 \atm i < |\queue|$ and $\from \queue i \satisfies \phi$.

  \item $\queue \satisfies \G \phi$ iff for all $i \in \NN$ such that $0 \atm i < |\queue|$, $\from \queue i \satisfies \phi$.

  \end{itemize}
  Satisfaction of Boolean combinations is defined as usual, e.g.\ $\queue
  \satisfies \lnot \phi$ iff $\queue \not\satisfies \phi$. No other pair
  $(\queue,\phi)$ satisfies $\queue \satisfies \phi$.
\end{DEF}
For instance, formula $\numb \event \atm 3$ is true exactly for queues
containing at most 3~$\event$'s, and formula $\G (\numb \event \atl 1)$ is
true of $\queue$ iff $\queue$ is empty or its final event (!) is
$\event$.\linebreak See \appref
B\Unless{\arxivVersion}{\ of~\cite{LWL19TR}} for more examples.

Algorithmically checking whether a concrete queue $\queue$ satisfies a QuTL
formula $\phi$ is straightforward, since $\queue$ is of fixed size and
straight-line. The situation is different with abstract queues. Our
motivation here is to declare that an abstract queue $\queueAbs$
\emph{violates} a formula $\phi$ if \emph{all its concretizations}
(\definitionref{concretizemap}) do: under this condition, if $\phi$ is an
invariant, we know $\queueAbs$ is not reachable. Equivalently:
\begin{DEF}[Abstract semantics of QuTL]
  \label{definition: QuTL abstract semantics}
  Abstract queue $\queueAbs$ \emphdef{satisfies} QuTL formula $\phi$,
  written $\queueAbs \satisfiesAbs \phi$, if some concretization of
  $\queueAbs$ satisfies $\phi$:
  \begin{equation}
    \label{equation: QuTL abstract semantics}
    \queueAbs \satisfiesAbs \phi \wbox{$:=$} \exists \queue \in \concretizeListMap[p](\queueAbs) : \queue \satisfies \phi \ .
  \end{equation}
\end{DEF}
For example, we have $\abstractqueue{bb}{ba} \satisfiesAbs[2] \G (a
\limplies \G \lnot b)$ since for instance $\mathit{bbba} \in
\concretizeListMap[2](\abstractqueue{bb}{ba})$ satisfies the formula. See
\appref B\Unless{\arxivVersion}{\ of~\cite{LWL19TR}} for more examples.

\subsection{Abstract QuTL Model Checking}
\label{section: Abstract QuTL Model Checking}

A QuTL \emph{constraint} is a QuTL formula without Boolean connectives. We
first describe how to model check against QuTL constraints, and come back
to Boolean connectives at the end of \sectionref{Abstract QuTL Model
Checking}.

Model checking an abstract queue $\queueAbs$ against a QuTL constraint
$\phi$, i.e.\ checking whether some concretization of $\queueAbs$ satisfies
$\phi$, can be reduced to a standard model checking problem over a labeled
transition system (LTS) $\lts = (\ltsStates,\ltsTranss, \ltsLabel)$ with
states $\ltsStates$, transitions $\ltsTranss$, and a labeling function
\mbox{$\func \ltsLabel \ltsStates {2^{\alphabet} \union
    \{\emptyword\}}$}. The LTS \emph{characterizes} the concretization
$\concretizeListMap[p](\queueAbs)$ of $\queueAbs$, as illustrated in
\figureref{LTS} using an example: the concretizations of $\queueAbs$ are
formed from the regular-expression traces generated by paths of
$\queueAbs$'s LTS that end in the double-circled green state.
\vskip-2pt
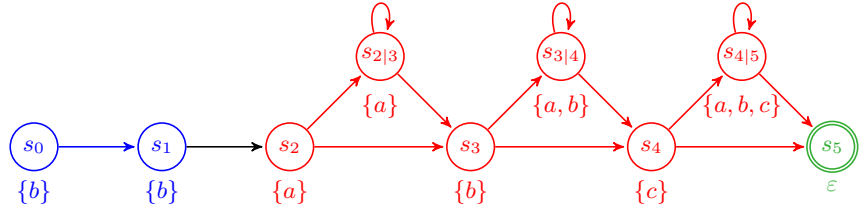
\begin{figure}[htbp]
  \begin{center}
    \begin{minipage}{.95\textwidth}
      \begin{center}
        \begin{tikzpicture}[->,>=stealth',shorten >=1pt,auto,node distance=17mm,semithick]
          \tikzstyle{every state}=[color = blue, draw,inner sep=0pt,minimum size=18pt]
          \node[state,color=blue]  (s0)                       {$s_0$};
          \node[state,color=blue]  (s1)  [      right of=s0]  {$s_1$};
          \node[state,color=red]   (s2)  [      right of=s1]  {$s_2$};
          \node[state,color=red]   (s23) [above right of=s2]  {$s_{2 \mid 3}$};
          \node[state,color=red]   (s3)  [below right of=s23] {$s_3$};
          \node[state,color=red]   (s34) [above right of=s3]  {$s_{3 \mid 4}$};
          \node[state,color=red]   (s4)  [below right of=s34] {$s_4$};
          \node[state,color=red]   (s45) [above right of=s4]  {$s_{4 \mid 5}$};
          \node[accepting,state,color=black!40!green!80] (s5)  [below right of=s45]  {$s_5$};

          \node[color=blue]              (l0)  [below =0.0 cm of s0]  {$\{b\}$};
          \node[color=blue]              (l1)  [below =0.0 cm of s1]  {$\{b\}$};
          \node[color=red]               (l2)  [below =0.0 cm of s2]  {$\{a\}$};
          \node[color=red]               (l23) [below =0.5 mm of s23] {$\{a\}$};
          \node[color=red]               (l3)  [below =0.0 cm of s3]  {$\{b\}$};
          \node[color=red]               (l34) [below =0.5 mm of s34] {$\{a,b\}$};
          \node[color=red]               (l4)  [below =0.0 cm of s4]  {$\{c\}$};
          \node[color=red]               (l5)  [below =0.5 mm of s45] {$\{a,b,c\}$};
          \node[color=black!40!green!60] (l6)  [below =0.0 cm of s5]  {$\emptyqueue$};

          \path (s0)  edge[color=blue]            node {} (s1)
                (s1)  edge                        node {} (s2)
                (s2)  edge [color=red]            node {} (s23)
                      edge [color=red]            node {} (s3)
                (s23) edge [loop above,color=red] node {} (s23)
                      edge [color=red]            node {} (s3)
                (s3)  edge [color=red]            node {} (s34)
                      edge [color=red]            node {} (s4)
                (s34) edge [loop above,color=red] node {} (s34)
                      edge [color=red]            node {} (s4)
                (s4)  edge [color=red]            node {} (s45)
                      edge [color=red]            node {} (s5)
                (s45) edge [loop above,color=red] node {} (s45)
                      edge [color=red]            node {} (s5);
        \end{tikzpicture}
      \end{center}
    \end{minipage}
  \end{center}
  \caption{LTS for $\queueAbs = \abstractqueue{bb}{abc}$ ($p=2$), with
    label sets written below each state. The blue and red parts encode the
    concretizations of the prefix and suffix of $\queueAbs$, resp.}
  \label{figure: LTS}
\end{figure}

The straightforward construction of the LTS $\lts$ is formalized in
\Ifthenelse{\arxivVersion}{\appendixref{Construction of the LTS for Abstract Queue}}{\appref{A.2} of~\cite{LWL19TR}}. Its size is
linear in~$|\queueAbs|$: $|\ltsStates| = p + 2 \times (|\queueAbs| - p) +
1$ and $|\ltsTranss| = p + 4 \times (|\queueAbs| - p)$.

We call a path through $\lts$ \emph{complete} if it ends in the right-most
state $s_z$ of~$\lts$ (green in \figureref{LTS}). The labeling function
extends to paths via $\ltsLabel(s_i \rightarrow \ldots \rightarrow s_j) =
L(s_i) \concat \ldots \concat L(s_j)$.
This gives rise to the following characterization of $\concretizeListMap[p](\queueAbs)$:
\newcounterset{LTSLEM}{\theDEF}
\begin{LEM}
  \label{lemma: concretization queueAbs is accepting paths}
  Given abstract queue $\queueAbs$ over alphabet $\alphabet$, let
  $\lts=(\ltsStates,\ltsTranss,\ltsLabel)$ be its~LTS.
  \begin{equation}
    \label{equation: concretization queueAbs is accepting paths}
    \concretizeListMap[p](\queueAbs) \wbox{$=$} \Union \ \{\lang(\ltsLabel(\ltsPath)) \in  2^{\closure{\alphabet}} \st \ \mbox{$\ltsPath$ is a complete path from $s_0$ in $\lts$}\} \ .
  \end{equation}
\end{LEM}
We say path $\ltsPath$ \emph{satisfies} $\phi$, written $\ltsPath
\satisfiesAbs \phi$, if there exists $\queue \in
\lang(\ltsLabel(\ltsPath))$ s.t.~$\queue \satisfies \phi$.
\begin{COR}
  \label{corollary: abstract satisfaction is finding satisfying path}
  Let $\queueAbs$ and $M$ as in \lemmaref{concretization queueAbs is
    accepting paths}, and $\phi$ a QuTL constraint. Then the following are
  equivalent.
  \begin{enumerate}

  \item $\queueAbs \satisfiesAbs \phi$.

  \item There exists a complete path $\ltsPath$ from $s_0$ in $\lts$ such
    that $\ltsPath \satisfiesAbs \phi$.

  \end{enumerate}
\end{COR}
\proof immediate from \definitionref{QuTL abstract semantics} and
\lemmaref{concretization queueAbs is accepting paths}.\eop

Given an abstract queue $\queueAbs$, its LTS $\lts$, and a QuTL constraint
$\phi$, our abstract queue model checking algorithm is based on
\corollaryref{abstract satisfaction is finding satisfying path}: we need to
find a complete path from $s_0$ in $\lts$ that satisfies $\phi$. This is
similar to standard model checking against existential temporal logics like
ECTL, with two particularities:

First, paths must be complete. This poses no difficulty, as completeness is
suffix-closed: a path ends in $s_z$ iff any suffix does. This implies that
temporal reductions on QuTL constraints work like in standard temporal
logics. For example: there exists a complete path $\ltsPath$ from $s_0$ in
$\lts$ such that $\ltsPath \satisfiesAbs \X \phi$ iff there exists a
complete path $\ltsPath'$ from some successor $s_1$ of $s_0$ such that
$\ltsPath' \satisfiesAbs \phi$.

Second, we have domain-specific atomic (non-temporal) propositions. These
are accommodated as follows, for an arbitrary start state $s \in S$:
\begin{description}

\item[$\exists \ltsPath: \ \mbox{$\ltsPath$ from $s$ complete and $\ltsPath \satisfiesAbs \event$}$ (for $\event \in \alphabet$):] \mbox{} \\
  this is true iff $\event \in L(s)$, as is immediate from the $\queue
  \satisfies \event$ case in \definitionref{QuTL concrete semantics}.

\item[$\exists \ltsPath: \ \mbox{$\ltsPath$ from $s$ complete and $\ltsPath \satisfiesAbs \numb \event > c$}$ (for $\event \in \alphabet, c \in \NN$):]
  this is true iff
  \begin{itemize}

  \item the number of states reachable from
    $s$ labeled $e$ is greater than $c$, \hfill \emphasize{or}

  \item there exists a state reachable from $s$ labeled with $e$ that has a
    self-loop.

  \end{itemize}
  The other relational expressions $\numb e \qrelop c$ are checked
   similarly.\eop

\end{description}

\paragraph*{Boolean connectives.} Let now $\phi$ be a full-fledged QuTL
formula. We first bring it into negation normal form, by pushing negations
inside, exploiting the usual dualities $\lnot\X = \X\lnot$, $\lnot\F = \G
\lnot$, and $\lnot\G = \F \lnot$. The subset $\qrelop \in
\{<,\atm,\atl,>\}$ of the queue relational expressions is semantically
closed under negation; ``$\lnot{=}$'' is replaced by ``$> \lor <$''. A path
$\ltsPath$ from $s$ satisfies $\lnot \event$ (for $\event \in \alphabet$)
iff $L(s) \not= \{e\}$: this condition states that either $L(s) =
\emptyword$, or there exists some label other than $e$ in $L(s)$, so the
\emph{existential} property $\lnot e$ holds.

Disjunctions are handled by distributing $\satisfiesAbs$ over them:
$\queueAbs \satisfiesAbs \phi_1 \lor \phi_2$ iff $\queueAbs \satisfiesAbs
\phi_1 \lor \queueAbs \satisfiesAbs \phi_2$. What remains are
conjunctions. The existential flavor of $\satisfiesAbs$ implies that
$\satisfiesAbs$ does \emph{not} distribute over them; see Ex.\ 13 in
\appref{B.1}\Unless{\arxivVersion}{\ of~\cite{LWL19TR}}. Suppose we ignore
this and replace a check of the form $\queueAbs \satisfiesAbs \phi_1 \land
\phi_2$ by the \emphasize{weaker} check $\queueAbs \satisfiesAbs \phi_1
\land \queueAbs \satisfiesAbs \phi_2$, which may produce false positives.
Now consider how we use these results: if $\queueAbs \satisfiesAbs \phi$
holds, we decide to \emph{keep} the state containing the abstract
queue. False positives during abstract model checks therefore may create
extra work, but do not introduce unsoundness. In summary, our abstract
model checking algorithm soundly approximates conjunctions, but remains
exact for the purely disjunctive fragment of QuTL.

\section{Empirical Evaluation}
\label{section: Empirical Evaluation}

We implemented the proposed approaches in \textsf{C}{\#} atop the bounded
model checker \Ptester~\cite{DGJQRZ13}, an analysis tool for \pfeature{P}
programs. \Ptester\ employs a bounded exploration strategy similar to
\zing~\cite{AQRRX04}. We denote by \Patp\ the implementation of
\algorithmref{Verifying P using abstracted observation sequences}, and by
\Pati\ the version with queue invariants (``\ourtool +
Invariants''). A~detailed introduction to tool design and implementation is available 
online~\cite{QubaWebsite}.

\paragraph{Experimental Goals.}
We evaluate the approaches against the following questions:
\begin{enumerate}[{\bf Q1.}]
\item Is \Patp\ effective: does it converge for many programs? for what
  values of $k$?

\item What is the impact of the QuTL invariant checking?

\end{enumerate}

\newcommand{\benchmark}[1]{\textbf{#1}}
\begin{table*}[t]
 \caption{Results: $\mathit{\#M}$: \#\pfeature{P} machines;
   $\mathit{Loc}$: \#lines of code;
   $\mathit{Safe?} = \Yes$: property holds;
   $p$: \emph{minimum} unabstracted prefix for required convergence;
   $\convergencePoint$: point of convergence or exposed bugs (-- means divergence);
   \emph{Time}: runtime (sec); \emph{Mem.}: memory usage~(Mb.).
 }
 \label{table: benchmarks}
 \centering
  \renewcommand{\arraystretch}{1.3}
\newcommand{\nores}{TO}
\resizebox{6.0cm}{!}{
  \noindent
  \begin{tabular}{@{}lccc@{}c@{}crrr@{}}
    \hline
    \multirow{2}{*}{ID/Program}
    & \multicolumn{3}{c}{Program Features}
    & \multirow{2}{*}{\ \ }
    & \multicolumn{4}{c}{\Patp}\\
    \cline{2-4}\cline{6-9}\noalign{\smallskip}
    & \multicolumn{1}{c}{~$\mathit{\#M}$~} 
    & \multicolumn{1}{r}{$\mathit{Loc}$} 
    & \multicolumn{1}{r}{$\mathit{Safe}?$} 
    &                                                                                 
    & \multicolumn{1}{c@{}}{~~$p$~~}                                                                                 
    & \multicolumn{1}{c@{}}{~~$\convergencePoint$~~} 
    & \multicolumn{1}{r}{~$Time$~}
    & \multicolumn{1}{r}{~$Mem.$~} \\
    \hline
    \multirow{1}{*}{\textsc{1/German-1\hspace{2mm}}}
    & 3 & 242 & $\Yes$ & & $4$ & $-$ & \nores & $-$ \\
    \hline
    \multirow{1}{*}{\textsc{2/German-2\hspace{2mm}}}
    & 4 & 244 & $\Yes$ & & $4$ & $-$ & \nores & $-$ \\
    \hline
    \multirow{1}{*}{\textsc{3/TokenRing-buggy \hspace{0mm}}}
    & 6 & 164 & $\No$ & & 0 & 2 & ~~241.44 & 35.96 \\
    \hline
    \multirow{1}{*}{\textsc{4/TokenRing-fixed \hspace{0mm}}}
    & 6 & 164 & $\Yes$ & & 0 & 4 &  1849.25 & 130.87 \\
    \hline
    \multirow{1}{*}{\textsc{5/FailureDetector\hspace{0mm}}}
    & 6 & 229 & $\Yes$ & & 0 & 4 & 183.99 & 12.38 \\
    \hline
    \multirow{1}{*}{\textsc{6/OSR \hspace{2mm}}}
    & 5 & 378 & $\Yes$ & & 0 & 5 &  77.92 & 44.86 \\
    \hline
    \multirow{1}{*}{\textsc{7/openWSN \hspace{2mm}}}
    & 6 & 294 & $\Yes$ & & 2 & 5 & 2574.25 & 376.29 \\
    \hline
  \end{tabular}}
\resizebox{6.0cm}{!}{
  \noindent
  \begin{tabular}{@{}lccc@{}c@{}crrr@{}}
    \hline
    \multirow{2}{*}{ID/Program}
    & \multicolumn{3}{c}{Program Features}
    & \multirow{2}{*}{\ \ }
    & \multicolumn{4}{c}{\Patp}\\
    \cline{2-4}\cline{6-9}\noalign{\smallskip}
    & \multicolumn{1}{c}{~$\mathit{\#M}$~} 
    & \multicolumn{1}{r}{$\mathit{Loc}$} 
    & \multicolumn{1}{r}{$\mathit{Safe}?$} 
    &                                                                                 
    & \multicolumn{1}{c@{}}{~~$p$~~}                                                                                 
    & \multicolumn{1}{c@{}}{~~$\convergencePoint$~~} 
    & \multicolumn{1}{r}{~$Time$~}
    & \multicolumn{1}{r}{~$Mem.$~} \\
    \hline
    \multirow{1}{*}{\textsc{8/Failover\hspace{2mm}}}
    & 4 & 132 & $\Yes$ & & 0 & 2 & 2.91 & 8.56\\
    \hline
    \multirow{1}{*}{\textsc{9/MaxInstances}}
    & 4 & 79 & $\Yes$ & & 0 & 3 & 0.14 & 0.56 \\
    \hline
    \multirow{1}{*}{\textsc{10/PingPong \hspace{2mm}}}
    & 2 & 76 & $\Yes$ & & 0 & 2 & 0.06  & 0.43 \\
    \hline
    \multirow{1}{*}{\textsc{11/BoundedAsync}}
    & 4 & 96 & $\Yes$ & & 0 & 5 & ~~203.39 & 29.32\\
    \hline
    \multirow[t]{3}{*}{\textsc{12/PingFlood \hspace{2mm}}}
    & 2 & 52 & $\Yes$ & & 4 & 5  & 0.11 & 0.43 \\
    \hline
    \multirow{1}{*}{\textsc{13/Elevator-buggy \hspace{0.5mm}}}
    & 4 & 270 & $\No$ & & 0 & 1 & 1.29 & 5.23\\
    \hline
    \multirow{1}{*}{\textsc{14/Elevator-fixed \hspace{0mm}}}
    & 4 & 271 & $\Yes$ & & 0 & 4 & 49.23 & 45.36\\
    \hline
  \end{tabular}}
\end{table*}

\paragraph{Experimental Setup.} We collected a set of \pfeature{P}
programs (available online~\cite{QubaWebsite}); most have been used in
previous publications:
\begin{description}
\item[1--5:] protocols implemented in \pfeature{P}: the
  German Cache Coherence protocol with different number of clients
  (\benchmark{1--2})~\cite{DGJQRZ13},
  a buggy version of a token ring protocol~\cite{DGJQRZ13}, and a fixed
  version (\benchmark{3--4}), and a failure detector protocol
  from~\cite{Pwebsite}~(\benchmark{5}).

\item[6--7:] two device drivers where OSR is used for testing USB
  devices~\cite{DGM14}.
  
\item[8--14:] miscellaneous: \benchmark {8--10}~\cite{Pwebsite}, \benchmark
  {11}~\cite{FHMP08}, \benchmark {12} is the example from
  \sectionref{Overview}, \benchmark {13--14} are the buggy and fixed
  versions of an Elevator controller~\cite{DGJQRZ13}.
\end{description} 
We conduct two types of experiments: (i) we run \Patp\ on each
benchmark to empirically answer {\bf Q1}; (ii) we run \Pati\ on the
examples which fail to verify in (i) to answer {\bf Q2}. All experiments
are performed on a 2.80 GHz Intel(R) Core(TM) i7-7600 machine with 8 GB
memory, running 64-bit Windows 10. The timeout is set to 3600sec (1h);
the memory limit to 4 GB.

\subsubsection{Results.} \tableref{benchmarks} shows that \Patp\ converges
on \emph{almost all} safe examples (and successfully exposes the bugs for
unsafe ones). Second,
in most cases, the $\convergencePoint$ where convergence was detected is
small, 5 or less. This is what enables the use of this technique in
practice: the exploration space grows fast with $k$, so early convergence
is critical. Note that $\convergencePoint$ is guaranteed to be the smallest
value for which the respective example converges. If convergent, the verification
succeeded fully automatically: the queue abstraction prefix parameter $p$
is incremented in a loop whenever the current value of $p$ caused a
spurious abstract state.

The \textsc{German} protocol does not converge in reasonable time.
In this case, we request minimal manual assistance from the designer. Our
tool inspects spurious abstract states, compares them to actually reached
abstract states, and suggests candidate invariants to exclude
them. We describe
the process of invariant discovery, and why and how they are easy to prove,
in~\cite{QubaWebsite}.

The following table shows the invariants that make the \textsc{German}
protocol converge, and the resulting times and memory consumption.\\[-5pt]

\begin{centering}
  \resizebox{12.2cm}{!}{
    \begin{tabular}{lccrrcll}
      \hline
      \multicolumn{1}{c@{}}{Program}
      & \multicolumn{1}{c@{}}{~~$p$~~}                                                                                 
      & \multicolumn{1}{c@{}}{~~$\convergencePoint$~~} 
      & \multicolumn{1}{r}{~$Time$~}
      & \multicolumn{1}{r}{~$Mem.$~}
      & \multicolumn{1}{r}{\ \ }  
      & \multicolumn{1}{c}{~Invariant~}\\
      \hline
      \multirow{1}{*}{\textsc{German-1\hspace{2mm}}} & 0 & 4 &  15.65  & 45.65 & \hspace{2mm}& \textsf{Server: $\numb{\mathit{req\_excl}} \atm 1 \land \numb{req\_share} \atm 1$} \\
      \cline{1-5}
      \multirow{1}{*}{\textsc{German-2\hspace{2mm}}} & 0 & 4 & \hspace{5mm}629.43 & \hspace{5mm}284.75  & & \textsf{Client: $\numb{\mathit{ask\_excl}} \atm 1 \land \numb{\mathit{ask\_share}} \atm 1$} \\
      \hline
    \end{tabular}
  }
\end{centering}

The invariant states that there is always at most one exclusive request
and at most one shared request
in the \textsf{Server} or \textsf{Client} machine's queue.

\paragraph{Performance Evaluation.} We finally consider the following
question: \emph{To perform full verification, how much overhead does
  \Patp\ incur compared to \Ptester?} We iteratively run \Ptester\ with a
queue bound from 1 up to $\convergencePoint$ (from \tableref{benchmarks}).

\noindent
\centerTwoOut{%
  \begin{minipage}{.35\textwidth}
    The figure on the right compares the running times of
    \ourtool\ and \Ptester. We observe that
    the difference is small, in all cases,
    suggesting that turning PTester into a full verifier comes with little
    extra cost. Therefore, as for improving \ourtool's scalability, the
    focus should
  \end{minipage}}{
  \begin{minipage}{.61\textwidth}
    \includegraphics[scale=0.33]{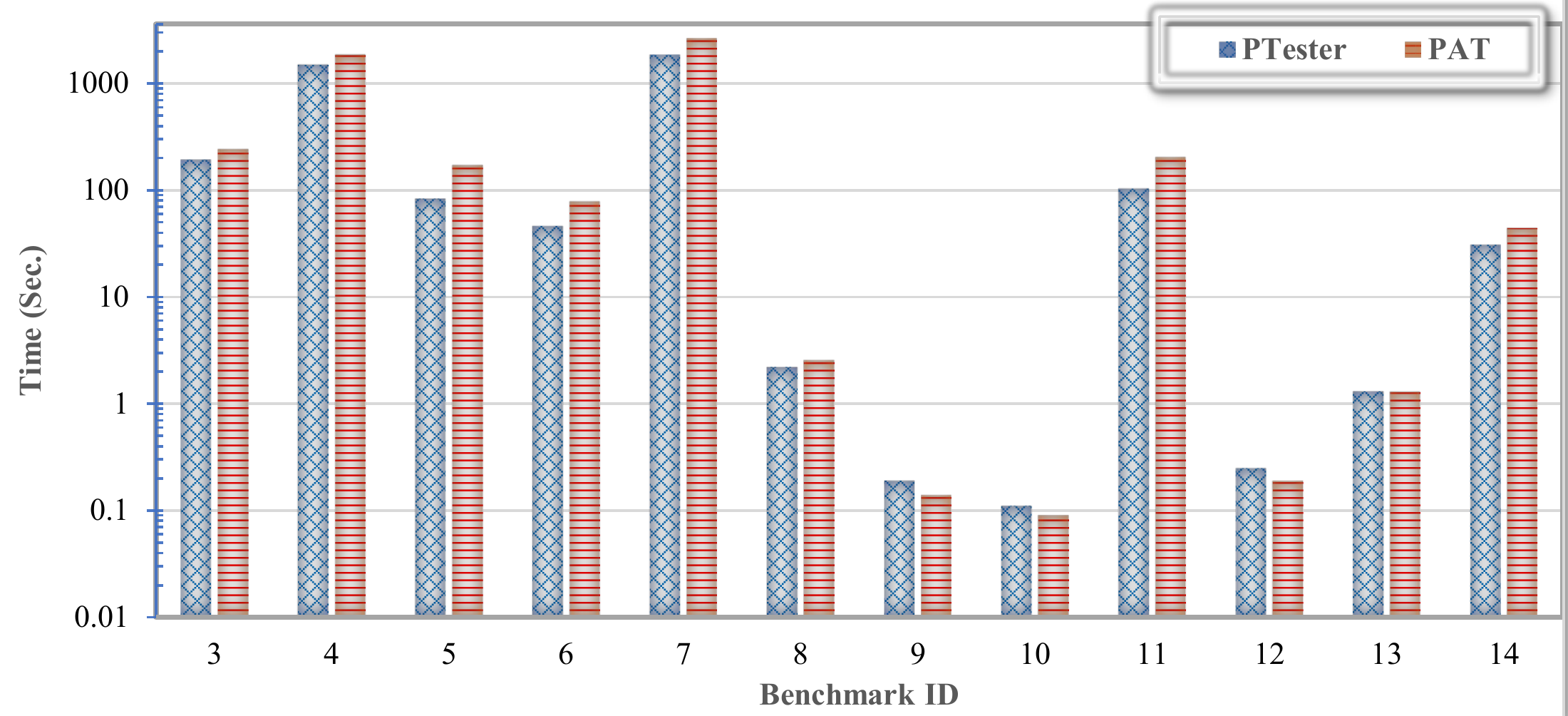}
  \end{minipage}
}

\noindent
be on the efficiency of the $\concreteReach[k]$ computation (\lineref{build
  Ak} in \algorithmref{Verifying P using abstracted observation
  sequences}). Techniques that\linebreak lend themselves here are
\emph{partial order reduction}~\cite{PSBK14,SRDK17} or \emph{symmetry
  reduction}~\cite{TD10}. Note that our proposed approach is orthogonal to
how these sets are computed.

\section{Related Work}

Automatic verification for asynchronous event-driven programs communicating
via unbounded FIFO queues is undecidable~\cite{BZ83}, even when the agents
are finite-state machines.
To sidestep the undecidability, various remedies are proposed. One is
to underapproximate program behaviors using various
bounding techniques; examples include
depth- \cite{G97} and context-bounded analysis~\cite{QR05,LMP09,LR09}, 
delay-bounding~\cite{EQR11}, bounded asynchrony \cite{FHMP08},
preemption-bounding~\cite{MQ07}, and
phase-bounded analysis~\cite{BE14,AAC13}.
It has been shown that most of these bounding techniques admit a decidable
model checking problem~\cite{QR05,LMP09,LR09} and thus have been
successfully used in practice for finding bugs.

Gall et.\ al proposed an abstract interpretation of FIFO queues
in terms of regular languages~\cite{GJJ06}. While our
works share some basic insights about taming queues, the differences are
fundamental: our abstract domain is \emph{finite}, guaranteeing convergence
of our sequence. In \cite{GJJ06} the abstract domain is infinite; they
propose a widening operator for fixpoint computation. More critically, we
use the abstract domain \emph{only} for convergence detection; the set of
reachable states returned is in the end exact. As a result, we can prove
and refute properties but may not terminate; \cite{GJJ06} is inexact and
cannot refute but always returns.

Several partial verification approaches for asynchronous message-passing
programs have been presented
recently~\cite{DGM14,BGKJ17,BEJQ18}. In~\cite{BGKJ17}, Bakst et
al.\ propose \emph{canonical sequentialization}, which avoids exploring all
interleavings by sequentializing concurrent programs.
Desai et al~\cite{DGM14} propose an alternative
way, namely by
prioritizing receive actions over send actions.
The approach is complete in the sense that it is able to construct
\emph{almost-synchronous invariants} that cover all reachable local states
and hence suffice to prove local assertions. Similarly, Bouajjani et
al.~\cite{BEJQ18} propose an iterative analysis that bounds send actions in
each interaction phase. It approaches the completeness by checking a
program's synchronizability under the bounds.
Similar to our work, the above three works are
sound but incomplete. An experimental comparison against the
  techniques reported in \cite{DGM14,BEJQ18} fails due to the
  unavailability of a tool that implements them. While tools implementing
these techniques are not available~\cite{DGM14,BEJQ18}, a comparison based
on what is reported in the papers suggests that our approach is competitive
in both performance and precision.


Our approach can be categorized as a \emph{cutoff}
detection technique~\cite{EK00,AHH13,FKP15,SRDK17}.
Cutoffs are, however, typically determined statically, often leaving them too large for
practical verification.
Aiming at minimal cutoffs, our work is closer in nature to earlier
\emph{dynamic} strategies \cite{KKW10,LW18},
which targeted different forms of concurrent programs. The
\emph{generator} technique proposed in \cite{LW18} is unlikely to work for
\pfeature P programs, due to the large local state space of machines.


\section{Conclusion}

We have presented a method to verify safety properties of asynchronous
event-driven programs of agents communicating via unbounded queues. Our
approach is sound but incomplete: it can both prove (or, by encountering
bugs, disprove) such properties but may not terminate. We empirically
evaluate our method on a collection of \pfeature{P} programs. Our
experimental results showcase our method can successfully prove the
correctness of programs; such proof is achieved with little extra resource
costs compared to plain state exploration. Future work includes an
extension to \pfeature P programs with other sources of unboundedness than
the queue length (e.g.\ messages with integer \emph{payloads}).


\paragraph{Acknowledgments.} We thank Dr.~Vijay D'Silva (Google, Inc.), for
enlightening discussions about partial abstract transformers.

%
\bibliographystyle{include/splncs04}
\bibliography{include/references}

\Ifthen{\arxivVersion}{
\clearpage
  
\appendix

\section*{Appendix}

\section{Proofs and Related Material}
\label{appendix: Proofs}

\subsection{Proof of Theorem \theoremref[]{convergence theorem}}
\label{appendix: Theorem convergence theorem}

Recall the definition of
$\batsymbol$:
\begin{equation}
  \label{equation: partial RkbarPrime (repeat)}
  \batsymbol = \{\abstractListMap[p](\globalstate') \st \exists \globalstate \in \concretizeListMap[p](\abstractReach[k]) : \globalstate \dequeuetranssymbol \globalstate'\} \ .
\end{equation}
\setcounter{DEF}{\theconvergenceTHE}
\begin{THE}
  If $\abstractReach[k] = \abstractReach[k-1]$ and $\batsymbol \subseteq
  \abstractReach[k]$, then for any $K \atl k$, $\abstractReach[K] = \abstractReach[k]$.
\end{THE}
\Proof: we need two lemmas.
\begin{LEM}
  \label{lemma: if de image contained then all images contained}
  Given $\abstractReach[k] = \abstractReach[k-1]$ and $\batsymbol \subseteq
  \abstractReach[k]$, we in fact have $(\abstractReach[k])' \subseteq
  \abstractReach[k]$, for the \emphasize{full} best abstract transformer
  image of $\abstractReach[k]$,
  \begin{equation}
    \label{equation: RkbarPrime}
    (\abstractReach[k])' = \{\abstractListMap[p](\globalstate') \st \exists \globalstate \in \concretizeListMap[p](\abstractReach[k]) : \globalstate \transsymbol \globalstate'\} \ .
  \end{equation}
\end{LEM}
\equationref{RkbarPrime} is identical to \equationref{partial RkbarPrime
  (repeat)}, except that it permits all CQS transitions in~$\transsymbol$,
not just those due to dequeue actions.

\Proof: we show $(\abstractReach[k])' \setminus \abstractReach[k] \subseteq
\batsymbol$. From this fact and the given $\batsymbol \subseteq
\abstractReach[k]$, we have $(\abstractReach[k])' \setminus
\abstractReach[k] \subseteq \abstractReach[k]$, which is equivalent to
$(\abstractReach[k])' \subseteq \abstractReach[k]$.

The above claim is equivalent to $(\abstractReach[k])' \setminus \batsymbol
\subseteq \abstractReach[k]$, which we now prove. Let $\bar \globalstate'
\in (\abstractReach[k])' \setminus \batsymbol$, i.e.\ $\bar \globalstate' =
\abstractListMap[p](\globalstate')$ for some $\globalstate,\globalstate',r$
such that $\globalstate = \concretizeListMap[p](\abstractListMap[p](r))$
(hence $\abstractListMap[p](s) = \abstractListMap[p](r)$), $r \in
\concreteReach[k]$, and $\globalstate \noqueuetranssymbol \globalstate'$ or
$\globalstate \enqueuetranssymbol \globalstate'$ (local or transmit
action). Since $\abstractListMap[p](s) = \abstractListMap[p](r)$, concrete
states $s$ and $r$ agree in all machines' local states, and in the prefixes
of length $p+1$ of all machines' queues. We now attempt to execute on $r$
the action that takes $s$ to $s'$. We distinguish what that action~is:
\begin{description}

\item[Case $\globalstate \noqueuetranssymbol \globalstate'$:] the
  internal action is executable on $r$ (it does not depend on any queue)
  and leads to a successor state $r'$ such that $\abstractListMap[p](r')
  = \abstractListMap[p](\globalstate')$. From $r \in \concreteReach[k]$ we
  have $r' \in \concreteReach[k]$ ($\concreteReach[k]$ = concrete
  reachability fixpoint), and thus $\bar \globalstate'
  = \abstractListMap[p](\globalstate')
  = \abstractListMap[p](r') \in \abstractListMap[p](\concreteReach[k])
  = \abstractReach[k]$.

\item[Case $\globalstate \enqueuetranssymbol \globalstate'$:] the transmit
  action is also executable on $r$, since it depends only on the local
  state (on which $s$ and $r$ agree), \emphasize{and, under bounded
    semantics, on the current size of the queue}: we must make sure the
  queue the action applies to is not full (of size $k$). This is the only
  place where we need the condition $\abstractReach[k] =
  \abstractReach[k-1]$: it ensures that in fact $r \in
  \concreteReach[k-1]$, hence all queues in $r$ have at most $k-1$
  events. The enqueue action can proceed; the rest of the proof is as
  in the previous case.

\end{description}
Observe that the above argument does not apply to dequeue actions: it is
not guaranteed that the states $\globalstate'$ and $r'$ obtained after
applying the dequeue are equivalent, i.e.\ that
$\abstractListMap[p](\globalstate') = \abstractListMap[p](r')$, because the
exact events in the abstracted parts of the queues are unknown and may
differ.
\begin{LEM}
  \label{lemma: claim main theorem}
  Given $\abstractReach[k] = \abstractReach[k-1]$ and $(\abstractReach[k])'
  \subseteq \abstractReach[k]$, we have for any $K \atl k$,
  $\abstractReach[K] = \abstractReach[k]$.
\end{LEM}
\Proof: by induction on $K$. The claim holds for $K=k$. Now suppose
$\abstractReach[K] = \abstractReach[k]$; we prove $\abstractReach[K+1] =
\abstractReach[k]$, which is equivalent to $\abstractReach[K+1] =
\abstractReach[K]$. Since $\abstractReach[K+1] \superseteq
\abstractReach[K]$, it suffices to show that $\abstractReach[K+1] \subseteq
\abstractReach[K]$.

To this end, let $a \in \abstractReach[K+1]$, i.e. $a =
\abstractListMap[p](s_a)$ for some $s_a \in \concreteReach[K+1]$. We show:
for all states $s$ along the path $\pi$ that reaches $s_a$,
$\abstractListMap[p](s) \in \abstractReach[K]$; call this claim~(*).
In~particular, then, $a = \abstractListMap[p](s_a) \in \abstractReach[K]$.

To show (*), we induct on the length of path $\pi$. The initial
state belongs to $\concreteReach[l]$ for every $l$, so its abstraction
belongs to $\abstractReach[K]$. Let now $s$ along $\pi$ be such that
$\abstractListMap[p](s) \in \abstractReach[K]$. Since $\abstractReach[K] =
\abstractReach[k]$, we have $s \in
\concretizeListMap[p](\abstractReach[k])$. By \equationref{RkbarPrime}, the
successor $s'$ of $s$ along $\pi$ satisfies $\abstractListMap[p](s') \in
(\abstractReach[k])' \subseteq \abstractReach[k] = \abstractReach[K]$.\eop

\Proof\ of \theoremref{convergence theorem}:

From $\abstractReach[k] = \abstractReach[k-1]$, $\batsymbol \subseteq
\abstractReach[k]$ and \lemmaref{if de image contained then all images
  contained} we conclude $(\abstractReach[k])' \subseteq
\abstractReach[k]$.

From $\abstractReach[k] = \abstractReach[k-1]$, $(\abstractReach[k])'
\subseteq \abstractReach[k]$ and \lemmaref{claim main theorem} we conclude
the claim in \theoremref{convergence theorem}.\eop

\subsection{Construction of the LTS for Abstract Queue $\queueAbs$}
\label{appendix: Construction of the LTS for Abstract Queue}

This construction is required for \lemmaref{concretization queueAbs is
  accepting paths} (\appendixref{Lemma concretization queueAbs is accepting
  paths}). We formalize it as follows. If $\queueAbs = \emptyqueue$, we let
$\ltsStates = \{s_0\}$, $\ltsTranss = \emptyset$, $\ltsLabel(s_0) =
\emptyword$. Otherwise, let $\queueAbs = \event[0] \ \ldots \ \event[p-1]
\ | \ \event[p] \ \ldots \ \event[z-1]$. We first define two separate LTS,
$\lts[p]$ and $\lts[s]$, for prefix and suffix, resp., and then conjoin
them to get $\lts$:
\setlength{\localLength}{3pt}
\begin{equation}
  \label{equation: LTS Mp and Ms}
  \begin{array}{lrcl}
                                                     & \ltsStates[p]            & = & \{ s_i : 0 \atm i < p \} \ , \\[\localLength]
    \lts[p] = (\ltsStates[p], \ltsTranss[p], \ltsLabel[p]):                           & \ltsTranss[p]            & = & \{ (s_i,s_{i+1}) : 0 \atm i < p-1 \} \ , \\[\localLength]
                                                     & \ltsLabel[p](s_i)       & = & \{ \event[i] \} \mbox{\ \ for $i$: $0 \atm i < p$} \\[5\localLength] 
                                                     & \ltsStates[s]            & = & \{ s_i,s_{i \mid i+1} : p \atm i < z \} \ \union \ \{s_z\} \ , \\[\localLength]
                                                     & \ltsTranss[s]            & = & \{  (s_i,s_{i|i+1}) , (s_{i|i+1},s_{i+1}), (s_i,s_{i+1}) , \\
                                                     &                &   & \ \, (s_{i|i+1},s_{i|i+1}) : p \atm i < z \} \\[\localLength]
    \raisetext{$\lts[s] = (\ltsStates[s], \ltsTranss[s], \ltsLabel[s]):$ \quad} & \ltsLabel[s](s_i)       & = & \{ \event[i] \}                  \mbox{\ \ for $i$: $p \atm i < z$} \ , \\[\localLength]
                                                     & \ltsLabel[s](s_{i|i+1}) & = & \{ \event[j] : p \atm j < i+1 \} \mbox{\ \ for $i$: $p \atm i < z$} \ , \\[\localLength]
                                                     & \ltsLabel[s](s_z)       & = & \emptyword \ .\footnotemark
  \end{array}
\end{equation}
\footnotetext{Note: if $p=z$ (empty suffix), $\lts[s]$ consists only of
  node $s_z$ (labeled~$\emptyword$) and no edges.} Now we define
$\lts=(\ltsStates,\ltsTranss,\ltsLabel)$ with
\begin{equation}
  \label{equation: LTS M}
  \ltsStates = \ltsStates[p] \union \ltsStates[s] \ , \quad \ltsTranss = \ltsTranss[p] \union \{(s_{p-1},s_p)\} \union \ltsTranss[s] \ , \quad \ltsLabel = \ltsLabel[p] \union \ltsLabel[s] \ .
\end{equation}

\newpage

\subsection{Proof of \lemmaref{concretization queueAbs is accepting paths}}
\label{appendix: Lemma concretization queueAbs is accepting paths}

\setcounter{DEF}{\theLTSLEM}
\begin{LEM}
  \label{lemma: appendix: concretization queueAbs is accepting paths}
  Given abstract queue $\queueAbs$ over alphabet $\alphabet$, let
  $M=(S,R,L)$ be its~LTS.
  \begin{equation}
    \label{equation: appendix: concretization queueAbs is accepting paths}
    \concretizeListMap[p](\queueAbs) \wbox{$=$} \Union \ \{\lang(\ltsLabel(\ltsPath)) \in 2^{\closure{\alphabet}} \st \ \mbox{$\ltsPath$ is a complete path from $s_0$ in $\lts$}\} \ .
  \end{equation}
\end{LEM}
\Proof: by induction on $|\queueAbs|$. If $\queueAbs = \emptyqueue$, then
$\concretizeListMap[p](\queueAbs) = \{\emptyqueue\}$. The only complete
path in $M$ is $\ltsPath=s_0$ with $\lang(L(s_0)) = \lang(\emptyword) =
\{\emptyword\}$.

\newcommand{\queueAbsAlt}{\overline T}
\newcommand{\queuePathAlt} t

Now suppose $\queueAbs = \queueAbsAlt.a$, i.e.\ $a$ is the final symbol of
$\queueAbs$, and let $M_{\queueAbsAlt}$ be $\queueAbsAlt$'s
LTS. $M_{\queueAbsAlt}$ is a sub-LTS (a ``prefix'' really) of $M$.
\renewcommand{\localCommand}{D\definitionref[]{concretizemap}}
\begin{itemize}

\item If $a$ is in the prefix of $\queueAbs$, i.e.\ the suffix of
  $\queueAbs$ is empty, we have
  \begin{eqnarray*}
    \concretizeListMap[p](\queueAbs) & \opNote{(\localCommand)} = & \concretizeListMap[p](\queueAbsAlt) \concat \{a\} \\
                                     & \opNote{(IH)}            = & \left( \Union \ \{\lang(L(\queuePathAlt)) \in 2^{\closure{\alphabet}} \st \ \mbox{$\queuePathAlt$ is a complete path from $s_0$ in $M_{\queueAbsAlt}$}\} \right) \concat \{a\}
  \end{eqnarray*}
  where ``\localCommand'' refers to \definitionref{concretizemap}, and
  ``IH'' denotes the induction hypothesis. Since $M$ equals
  $M_{\queueAbsAlt}$ extended by an edge to an $a$-labeled state, with a
  unique complete path, \equationref{appendix: concretization queueAbs is
    accepting paths} follows.

\item If $a$ is in the suffix of $\queueAbs$, then let $\alphabet_s$ be the
  set of \emph{all} suffix symbols of $\queueAbs$ (in particular, $a \in
  \alphabet_s$).
  \begin{eqnarray*}
    \concretizeListMap[p](\queueAbs) & \opNote{(\localCommand)} = & \concretizeListMap[p](\queueAbsAlt) \concat \{a\} \concat {\alphabet_s}^{*} \\
                                     & \opNote{(IH)}            = & \left( \Union \ \{\lang(L(\queuePathAlt)) \in 2^{\closure{\alphabet}} \st \ \mbox{$\queuePathAlt$ is a complete path from $s_0$ in $M_{\queueAbsAlt}$}\} \right) \\
    & & \format\hspace*{70mm} \mbox{} \concat \{a\} \concat {\alphabet_s}^{*} \ .
  \end{eqnarray*}
  By the LTS construction in \appendixref{Construction of the LTS for
    Abstract Queue}, $M$ equals $M_{\queueAbsAlt}$ with an $a$-labeled
  state and a $\alphabet_s$-labeled state inserted before the right-most
  state, from which \equationref{appendix: concretization queueAbs is
    accepting paths} follows.\eop

\end{itemize}

\section{Additional Material}
\label{appendix: Additional Material}

\subsection{Examples for \sectionref{Abstract Queue Invariant Checking}}
\label{appendix: Examples for section Abstract Queue Invariant Checking}

\begin{EXA}
  \label{example: QuTL formulas}
  Here are some examples of QuTL formulas and their intuitive meanings. Let
  $\queue$ be a queue; $a \limplies b$ abbreviates $\lnot a \lor b$.
  \begin{quote}
    \begin{tabular}{l|l}
      Satisfaction relation & Meaning \\
      \hline
      $\queue \satisfies \numb \event \atm 3$                         & $\queue$ contains at most 3 $\event$'s. \\

      $\queue \satisfies \G (\event[1] \limplies \G \lnot \event[2])$ & In $\queue$, $\event[1]$ is never (eventually) followed by $\event[2]$. \\

      $\queue \satisfies \F (\numb \event < 2)$                       & $\queue$ is non-empty. \\

      $\queue \satisfies \G (\numb \event \atl 2)$                    & $\queue$ is empty. \\

      $\queue \satisfies \G (\numb \event \atl 1)$                    & $\queue$ is empty or its tail event is an $\event$.

    \end{tabular}
  \end{quote}
\end{EXA}

\begin{EXA}
  \label{example: QuTL abstract satisfaction}
  Here are some examples of abstract (non-)satisfaction. As indicated,
  these assume $p=2$, i.e.\ the first two queue events (if any) are
  unabstracted. Again, $a \limplies b$ abbreviates $\lnot a \lor b$.
  \[
    \abstractqueue{bb}{ba} \wbox{$\satisfiesAbs[2]$} \G (a \limplies \G \lnot b)
  \]
  since there is a concretization, for instance $\mathit{bbba}$, that
  satisfies the formula.
  \[
    \abstractqueue{ac}{b} \wbox{$\not\satisfiesAbs[2]$} \G (a \limplies \X b)
  \]
  since the violation is caused by the prefix of queue ($\mathit{ac}$), so
  \emph{all} concretizations violate this formula.
\end{EXA}

\begin{EXA}
  \label{example: satisfiesAbs not distribute over conjunction}
  Relation $\satisfiesAbs$ does not distribute over conjunction,
  i.e.\ $\queueAbs \satisfiesAbs \phi_1 \land \phi_2$ is not equivalent to
  $\queueAbs \satisfiesAbs \phi_1 \land \queueAbs \satisfiesAbs \phi_2$:
  Let $\queueAbs = \abstractqueue{a}{ab}$ and $\phi = \psi \land \neg \psi$
  where $\psi = (\numb a \atl 3)$. Clearly, $\queueAbs \not\satisfiesAbs[1]
  \phi$ since $\phi \equiv \false$. However, $\queueAbs \satisfiesAbs[1]
  \psi$ and $\queueAbs \satisfiesAbs[1] \neg\psi$ both hold, witnessed by
  two distinct concrete queues $\queue_{\psi} = \mathit{aaab}$ and
  $\queue_{\lnot \psi} = \mathit{aab}$.
\end{EXA}
}

\end{document}